\documentclass{emulateapj-rtx4}
\usepackage{apjfonts}
\usepackage{amssymb}
\usepackage{natbib}
 \usepackage{color}

\usepackage{graphicx,mathrsfs,amsmath}

\shorttitle{The radial and azimuthal profiles of Mg II absorption around galaxies}
\shortauthors{Bordoloi et al.}

\begin{document}
\title{THE RADIAL AND AZIMUTHAL PROFILES OF M\lowercase{g} II ABSORPTION AROUND 0.5 < \lowercase{z} < 0.9 \lowercase{z}COSMOS GALAXIES OF DIFFERENT COLORS, MASSES AND ENVIRONMENTS\altaffilmark{+}}
\author{R. Bordoloi\altaffilmark{1}, 
S. J. Lilly\altaffilmark{1}
  , C. Knobel\altaffilmark{1}
  , M. Bolzonella\altaffilmark{2}
  , P. Kampczyk\altaffilmark{1}
  , C.M. Carollo\altaffilmark{1} 
  , A. Iovino\altaffilmark{3}
  , E. Zucca\altaffilmark{2}
  ,T. Contini\altaffilmark{4,5}
  , J. -P. Kneib\altaffilmark{6}
  , O. Le Fevre\altaffilmark{6}
  , V. Mainieri\altaffilmark{7}
  , A. Renzini\altaffilmark{8}
  , M. Scodeggio\altaffilmark{9}
  , G. Zamorani\altaffilmark{2}
  , I. Balestra\altaffilmark{10}
  , S. Bardelli\altaffilmark{2}
  , A. Bongiorno\altaffilmark{10}
  , K. Caputi\altaffilmark{12}
  , O. Cucciati\altaffilmark{11}
  , S. de la Torre\altaffilmark{12}
  , L. de Ravel\altaffilmark{12}
  , B. Garilli\altaffilmark{9,6}
  , K. Kova\v{c}\altaffilmark{1,23}
  , F. Lamareille\altaffilmark{4,5}
  , J. -F. Le Borgne\altaffilmark{4,5}
  , V. Le Brun\altaffilmark{6}
  , C. Maier\altaffilmark{1}
  , M. Mignoli\altaffilmark{2}
  , R. Pello\altaffilmark{4,5}
  , Y. Peng\altaffilmark{1}
  , E. Perez Montero\altaffilmark{4,5,13}
  , V. Presotto\altaffilmark{3}
  , C. Scarlata\altaffilmark{22}
  , J. Silverman\altaffilmark{14}
  , M. Tanaka\altaffilmark{14}
  , L. Tasca\altaffilmark{6}
  , L. Tresse\altaffilmark{6}
  , D. Vergani\altaffilmark{2}
  , L. Barnes\altaffilmark{1}
  , A. Cappi\altaffilmark{2}
  , A. Cimatti\altaffilmark{15}
  , G. Coppa\altaffilmark{10}
  , C. Diener\altaffilmark{1}
  , P. Franzetti\altaffilmark{9}
  , A. Koekemoer\altaffilmark{16}
  , C. L\'{o}pez-Sanjuan\altaffilmark{6}
  , H.J. McCracken\altaffilmark{17}
  , M. Moresco\altaffilmark{18}
  , P. Nair\altaffilmark{2}
  , P. Oesch\altaffilmark{1,19}
  , L. Pozzetti\altaffilmark{2}
 \& N. Welikala\altaffilmark{21}
 }

\email{rongmonb@phys.ethz.ch}
 
 \altaffiltext{+}{based on observations undertaken at the European Southern Observatory (ESO) Very Large Telescope (VLT) under Large Program 175.A-0839}
\altaffiltext{1}{Institute for Astronomy, ETH Z\"{u}rich, Wolfgang-Pauli-Strasse 27, 8093, Z\"{u}rich, Switzerland}
\altaffiltext{2}{INAF Osservatorio Astronomico di Bologna, Bologna, Italy}
\altaffiltext{3}{INAF Osservatorio Astronomico di Brera, Milan, Italy}
\altaffiltext{4}{Institut de Recherche en Astrophysique et Plan\'{e}tologie, CNRS, 14, avenue Edouard Belin, F-31400 Toulouse, France }
\altaffiltext{5}{IRAP, Universit\'{e} de Toulouse, UPS-OMP, Toulouse, France}
\altaffiltext{6}{Laboratoire d'Astrophysique de Marseille, CNRS/Aix-Marseille Universit\'{e}, 38 rue Fr\'{e}d\'{e}ric Joliot-Curie, 13388, Marseille cedex 13, France}
\altaffiltext{7}{European Southern Observatory, Garching, Germany}
\altaffiltext{8}{Dipartimento di Astronomia, Universita di Padova, Padova, Italy}
\altaffiltext{9}{INAF - IASF Milano, Milan, Italy}
\altaffiltext{10}{Max Planck Institut f\"{u}r Extraterrestrische Physik, Garching, Germany}
\altaffiltext{11}{INAF-Osservatorio Astronomico di Trieste, Trieste, Italy}
\altaffiltext{12}{SUPA, The University of Edinburgh, Royal Observatory, Blackford Hill, Edinburgh EH9 1BD, UK}
\altaffiltext{13}{ Instituto de Astrof\'{i}sica de Andaluc\'{i}a, CSIC, Apartado de correos 3004, 18080 Granada, Spain}
\altaffiltext{14}{Institute for the Physics and Mathematics of the Universe (IPMU), University of Tokyo, Kashiwanoha 5-1-5, Kashiwa-shi, Chiba
277-8568, Japan}
\altaffiltext{15}{Dipartimento di Astronomia, Universit\`{a} degli Studi di Bologna, Bologna, Italy}
\altaffiltext{16}{Space Telescope Science Institute, Baltimore, Maryland 21218, USA}
\altaffiltext{17}{Institut d'Astrophysique de Paris, UMR7095 CNRS, Universit\'{e} Pierre \& Marie Curie, 75014 Paris, France}
\altaffiltext{18}{Dipartimento di Astronomia, Universit\`{a} degli Studi di Bologna, Bologna, Italy}
\altaffiltext{19}{UCO/Lick Observatory, Department of Astronomy and Astrophysics, University of California, Santa Cruz, CA 95064}
\altaffiltext{21}{Institut d'Astrophysique Spatiale, Batiment 121, CNRS \& Univ. Paris Sud XI, 91405 Orsay Cedex, France}
\altaffiltext{22}{School of Physics and Astronomy, University of Minnesota, Minneapolis, MN 55455}
\altaffiltext{23}{Max-Planck-Institut f\"{u}r Astrophysik, Garching, Germany}

\begin{abstract}
We map the radial and azimuthal distribution of Mg II gas within $\sim$ 200 kpc (physical) of $\sim$ 4000 galaxies at redshifts $0.5 < z < 0.9$ using co-added spectra of more than 5000 background galaxies at $z > 1$.  We investigate the variation of Mg II rest frame equivalent width as a function of the radial impact parameter for different subsets of foreground galaxies selected in terms of their rest-frame colors and masses.  Blue galaxies have a significantly higher average Mg II equivalent width at close galacto-centric radii as compared to the red galaxies. Amongst the blue galaxies, there is a correlation between Mg II equivalent width and galactic stellar mass of the host galaxy. We also find that the distribution of Mg II absorption around group galaxies is more extended than that for non-group galaxies, and that groups as a whole have more extended radial profiles than individual galaxies. Interestingly, these effects can be satisfactorily modelled by a simple superposition of the absorption profiles of individual member galaxies, assuming that these are the same as those of non-group galaxies, suggesting that the group environment may not significantly enhance or diminish the Mg II absorption of individual galaxies. We show that there is a strong azimuthal dependence of the Mg II absorption within 50 kpc of inclined disk-dominated galaxies, indicating the presence of a strongly bipolar outflow aligned along the disk rotation axis.  There is no significant dependence of Mg II absorption on the apparent inclination angle of disk-dominated galaxies. 

\end{abstract}

\section{Introduction}

Metal absorption lines, such as those of the Mg II $\lambda \lambda$ 2796, 2803 doublet, in the spectra of background sources such as quasars, provide an important tracer of enriched gas that is otherwise very hard to detect. Mg II absorption originates in photo-ionized gas at temperatures around T$\sim 10^{4}$ K  \citep{Bergeron1986, Charlton2003} with neutral Hydrogen column densities of  $N(HI) \approx 10^{16}-10^{22} \; cm^2$ \citep{Churchill2000, rigby2002, Rao2006}.  Optical spectrographs can detect the Mg II doublet over a broad redshift range of $z\sim 0.3-2.5$ and over the years there have been a large number of studies using QSO absorption lines to characterize the statistical properties of Mg II absorbers.

These works have studied the distribution of column densities, the redshift evolution of number densities and the kinematic signatures (see e.g. \citealt{Lanzetta1987, Sargent1988, Petitjean1990, Steidel1992, Charlton1998, Nestor2005,Prochter2006}) and have also extended to the detection of magnetic fields associated with them \citep{Bernet2008}. The association of strong Mg II absorption with normal, bright, field galaxies is by now well established (e.g. \citealt{Churchill2005a} and references herein). The Mg II gas around galaxies is traced out to $\sim 100 $ kpc with absorber covering fractions of 50-80$\%$ \citep{Chen2010a}.  The profile of the Mg II haloes around galaxies have also been studied quite extensively. \cite{Steidel1995} showed that their sizes scale weakly with galaxy luminosity and inferred that these halos have quite sharp boundaries. \cite{Churchill2005} searched for correlations between gas kinematics and galaxy orientation and found no significant correlation between the two for their small sample of absorbers. \cite{Kacprzak2008} suggested that covering fractions might be less that unity and that Mg II halos around galaxies are patchy and might have non-symmetric geometric distribution.

However despite such observational progress, the origin of the absorbing gas in the halos of galaxies is still widely debated. It has been postulated that strong Mg II systems originate due to cool gas entrained in outflows from star-forming galaxies \citep{Bouche2006,bouche2008,Menard2009, Nestor2010}, which is also supported by the blueshifts of Mg II absorption in the spectra of individual galaxies \citep{Wiener2009, Rubin2010}.  Others have suggested in falling gas as the source \citep{Chen2010a, 2008C&T,Kacprzak2010a}, or a combination of both inflows and outflows \citep{Chen2010b, Chelouche2010}. 

It has also been suggested that group galaxies may have different radial profiles in Mg II absorption compared with isolated galaxies \citep{Chen2010a}. A number of mechanisms for this have been suggested, such as tidal tails and streams etc. \citep{Bowen1995,Churchill&Charlton99,Kacprzak2010a,Kacprazak2010b}.  Recent observational studies have utilized close galaxy pairs and galaxy groups to study Mg II absorption strength around such systems \citep{Chen2010a, Nestor2007,Kacprzak2010a,Kacprazak2010b} but these studies have generally been limited to small samples of absorbers. 

Most of the above investigations have utilized the spectra of bright quasars for which high resolution, high S/N spectra can be readily obtained. More recently, it has become practical to use the spectra of background galaxies that have been observed in large scale redshift surveys \citep{Steidel2010}. The spectra of these are sufficiently noisy such that it is usually necessary to stack the spectra of many background galaxies at a given impact parameter $b$, so as to be able to detect the absorption signal.  This approach differs from that of using quasar spectra in a number of significant ways

\begin{figure*}
\begin{center}
    \includegraphics[angle=0,width=12cm,height=10cm]{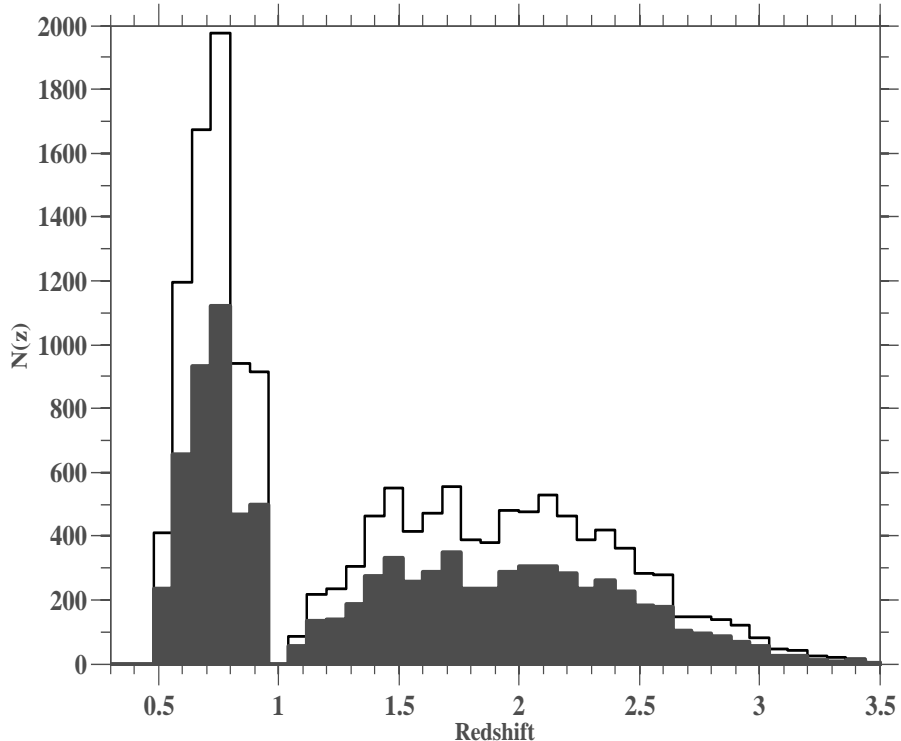}
\end{center}
\caption{The N(z) distributions of the foreground ($0.5 < z_{spec} < 0.9$) and background ($z_{phot} > 1$) galaxy samples. The open histograms represent the distribution of all galaxies and the shaded areas give the distribution of galaxies used for this analysis, i.e. those forming foreground-background pairs with projected separations of less than 200 kpc at the foreground redshift.}

\label{N_z_distribution}
\end{figure*}

\begin{itemize}

\item  the individual lines of sight probe regions that are of order 10 kpc across, or more, as compared with the 1 pc sightlines of quasar spectra.
\item Furthermore, the stacked spectra for a given range of $b$, yield a measure of the average absorption over an annular region around the foreground systems encompassing an area of several thousand $\rm{kpc^{2}}$, i.e. about $10^9$ times larger in area as compared to a QSO sight-line. 
\item  This average spectrum will integrate the contribution of individual absorption systems that have too low equivalent width to be detected in even high S/N quasar spectra.
\item  Some of the difficulties that are encountered in correctly associating foreground galaxies with individual absorption systems (e.g. due to the glare of the bright quasar or limitations in the depth of the associated galaxy follow-up) are mitigated by using the spectra of background galaxies around a well defined set of foreground galaxies, rather than starting with a quasar spectrum and and trying to identify nearby foreground galaxies responsible for the absorption.

\end{itemize}

This approach is inherently statistical, since generally the absorption around multiple foreground galaxies must be co-added.  Against this, the number of spectra that are potentially available in large redshift surveys can be high enough to allow the statistical study of different sub-samples of foreground populations. This approach is orthogonal to most of the previous studies of Mg II absorption systems, in the sense that those studies select the samples of galaxies from detected Mg II absorption, comparing these with the general galaxy population.  Our own approach takes a well defined population of galaxies and then measures the absorbing gas around them. It should be noted that because we are summing (or averaging), the spectra across the spatial rather than the spectral domain, the derived equivalent widths will be correct, independent of the degree of saturation of the absorption along any given line of sight.    This is different from the case of absorption components which overlap spectrally along a given line of sight where saturation effects must be carefully considered.  Of course, conversion of our mean equivalent width to a mean column density along the annulus would need to consider the effects of saturation, but this will not be attempted in this paper.

The aim of this paper is to apply this approach to the available galaxy spectra in the zCOSMOS survey \citep{Lilly2007}. This is a large redshift survey that has been undertaken in the COSMOS field \citep{Scoville2007}. In particular we use the blue spectra of galaxies in zCOSMOS-deep, selected to have $z_{phot} > 1$ to probe the Mg II absorption around foreground galaxies that have secure spectroscopic redshifts from zCOSMOS-bright in the $0.5 < z < 0.9$ range.  

There are two major advantages of doing this study in the COSMOS field. First, the high density of spectroscopic redshifts yields good environmental information for the foreground galaxies in the form of a high fidelity group catalogue (\cite{Knobel_2009}, Knobel et al in prep). Second, high resolution HST/ACS images \citep{Koekemoer2007} are available for all galaxies, enabling us to study the azimuthal dependence of Mg II absorption relative to the disk axis for disk dominated galaxies.

This paper is organized as follows. In Section 2 we first present the spectroscopic datasets that are used, and describe the selection criteria and the derivation of the final sample. In Section 3 we construct the radial profile of Mg II absorption around foreground galaxies at $0.5 < z < 0.9$ and examine the dependence of this on the color, stellar mass and group environment of the galaxies.  In Section 3.1 we show directly that there is a clear mass dependence in the strength of Mg II absorption around blue galaxies, and that, at a given mass, the absorption around blue galaxies is much stronger than around red ones.  In Section 3.2, we show that the radial dependence of Mg II absorption around group galaxies is considerably more extended than around isolated galaxies. We also construct the absorption profile around the group centres, and around the most massive members of the groups, and show that this is also more extended than around isolated galaxies. We show that these effects can however be fully explained by a simple model which superposes the normal absorption profiles of the individual members. In Section 3.3 we study the azimuthal dependence of Mg II absorption around inclined disk galaxies and show that the absorption along the projected rotation axes of the disks is about three times as strong as that in perpendicular directions. Finally, in Section 3.4 we investigate the dependence of Mg II absorption on the apparent inclination angles of disk galaxies. 

Throughout this paper, we use a concordance cosmology with $\Omega_{m} = 0.25$, $\Omega_{\Lambda} = 0.75$ and $\rm{H_{0} = 70\; km\; s^{-1}\; Mpc^{-1}}$. Unless stated otherwise all magnitudes are given in the AB system. \textit{Because we are working with very low resolution spectra, all equivalent widths are quoted integrating over both components of the Mg II $\lambda \lambda$ 2796, 2803 doublet. They are therefore approximately twice as large as the equivalent widths of the individual components.}	

\section{Spectroscopic data}
\subsection {The zCOSMOS Redshift Survey}
The zCOSMOS survey \citep{Lilly2007} is a survey in the 2 square degree COSMOS field \citep{Scoville2007}. The zCOSMOS survey is carried out with the VIMOS spectrograph on the ESO UT3 8-m VLT. The survey itself is divided into two major components : 

1)  zCOSMOS-bright, which consists of spectra from approximately 20,000 flux limited $I_{AB} \leq 22.5$ galaxies over the full $\rm{2}$ square degree COSMOS field. At this flux limit, the majority of the observed galaxies have redshifts in the range of $0 < z < 1.4$. Observations in zCOSMOS-bright were obtained with the MR grism using 1 arcsec slits, yielding a spectral resolution of $\rm{R \sim 600}$ at $\rm{2.5 }$ {\AA} $\rm{ pixel^{-1}}$.  The average accuracy of individual redshifts has been demonstrated to be $\rm{110\; km s^{-1}}$ \citep{Lilly2009_Article}.  Stellar masses for these galaxies have been estimated by spectral energy distribution (SED) fitting, using the $Hyperzmass$ code, a modified version of the $photo-z$ code $Hyperz$ \citep{Hyperz}. We refer the reader to \cite{Bolzonella2010} for a detailed technical description of this mass estimation. The absolute magnitudes used in this study were computed as in \cite{Zucca2009}. 

2)  zCOSMOS-deep, targets the central 1 square degree of the field and consists of approximately 10,000 spectra of B $<$ 25.25 color-selected galaxies which are expected to predominantly lie in the redshift range of $1.4 < z < 3$ (\cite{Lilly2007}, Lilly et al in preparation). The spectra were obtained using the LR blue grism with a resolution R $ \sim $ 200 at $\rm{5.3}$ {\AA} $\rm{ pixel^{-1}}$. The spectra cover the wavelength range from 3500 {\AA} to 7000 \AA . For each of the objects observed spectroscopically, photometric redshifts are available from the excellent COSMOS photometry using a number of codes (e.g. \citealt{2009ApJ...690.1236I}).  The photo-z accuracy for these objects for which reliable spectroscopic redshifts are available is $ \sigma_{\Delta z}\sim 0.0381(1+z)$. Further details about these two parts of zCOSMOS are given in \cite{Lilly2007, Lilly2009_Article} and Lilly et al in prep.

\begin{figure*}
{\includegraphics[angle=0,width=9.3cm,height=8.5cm]{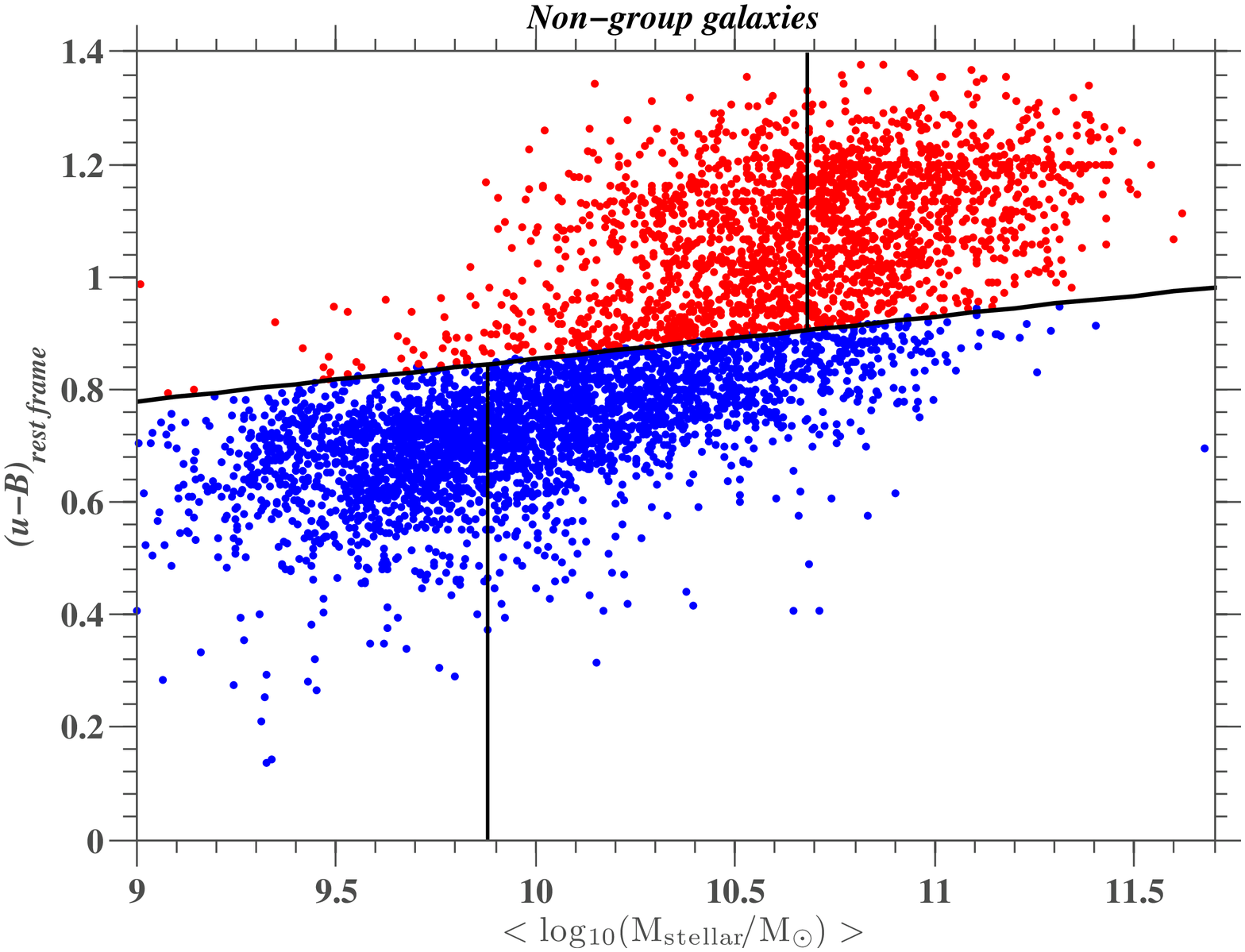}}
{\includegraphics[angle=0,width=9.3cm,height=8.5cm]{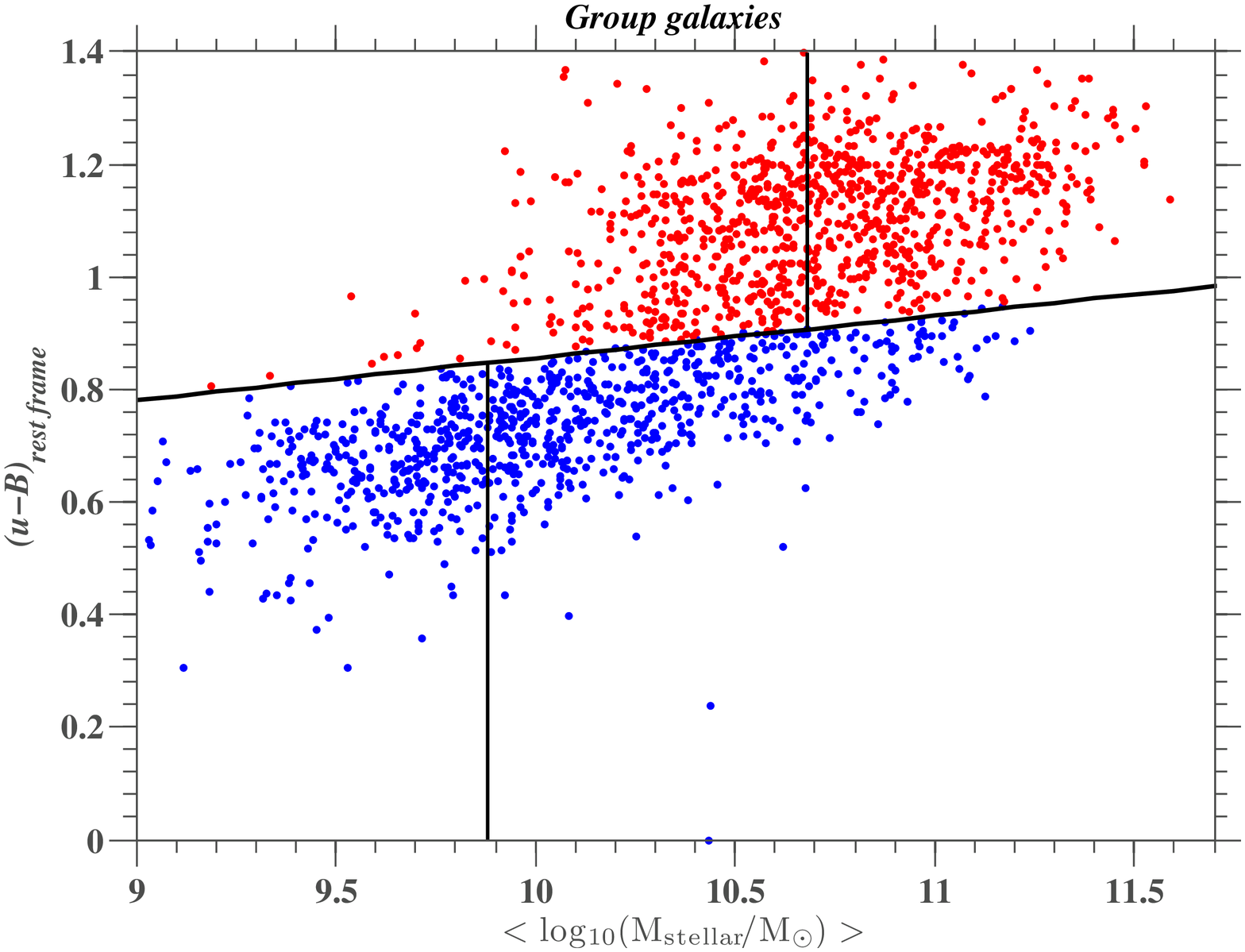}}

\caption{Foreground non-group and group galaxies divided into blue and red sub-samples as shown by the horizontal line computed from  equation:-\ref{mass_color_division}. The left panel gives the selected non-group sample and the right panel gives the selected group sample. For both the panels the red points are for red galaxies and the blue points are for blue galaxies. The vertical lines on either panel are the boundaries chosen to divide the sample in mass as described in Section 3.2}
\label{galaxy_color_mass_diagram}
\end{figure*}

\subsection{Foreground and background galaxy selection}

The first step in the analysis is to identify pairs of galaxies consisting of one foreground galaxy from zCOSMOS-bright, selected to lie at $0.5 < z_{spec} < 0.9$, and one background galaxy from zCOSMOS-deep whose sightline passes within a specified projected radius (or impact parameter) $b$ from the foreground galaxy.  The foreground redshift range of $0.5 < z_{spec} < 0.9$ is chosen so that the Mg II 2799 absorption at the foreground redshift lies within the spectral range of the zCOSMOS-deep spectra.  For the foreground galaxies we only use galaxies with high confidence redshift measurements i.e. those with redshift Confidence Classes of 4.x, 3.x, 2.5 and 1.5. As shown in \cite{Lilly2009_Article} these are 99\% reliable.  Based on the photo-z estimates of the other galaxies, this secure sample is, in this redshift range, about 95\% complete. 

The background sample is selected from the zCOSMOS-deep survey. Because we do not need to know the exact redshift of the background object, provided only that we know it truly lies well behind the foreground one, we can use the spectra of background galaxies even if a confident redshift has not been secured. We therefore select the background galaxies using their $photometric$ redshifts to have $\rm{z_{phot} > 1.0}$. 

Figure \ref{N_z_distribution} shows the redshift distributions of the foreground galaxies and the background galaxies, both of the full samples and of those galaxies that are used in this analysis because they form suitably close projected pairs.  The total number of foreground galaxies between $0.5 < z_{spec} < 0.9$ is 7110.  Of these, 3908 have at least one background galaxy projected within $b \leq 200$ kpc. These background galaxies comprise 5237 of the total potential background sample of 8529  galaxies. As expected, the samples that can be used for this study, are a random set of the parent samples (Figure \ref{N_z_distribution}).

\begin{figure*}
\begin{center}
{\includegraphics[angle=0,width=12cm,height=10cm]{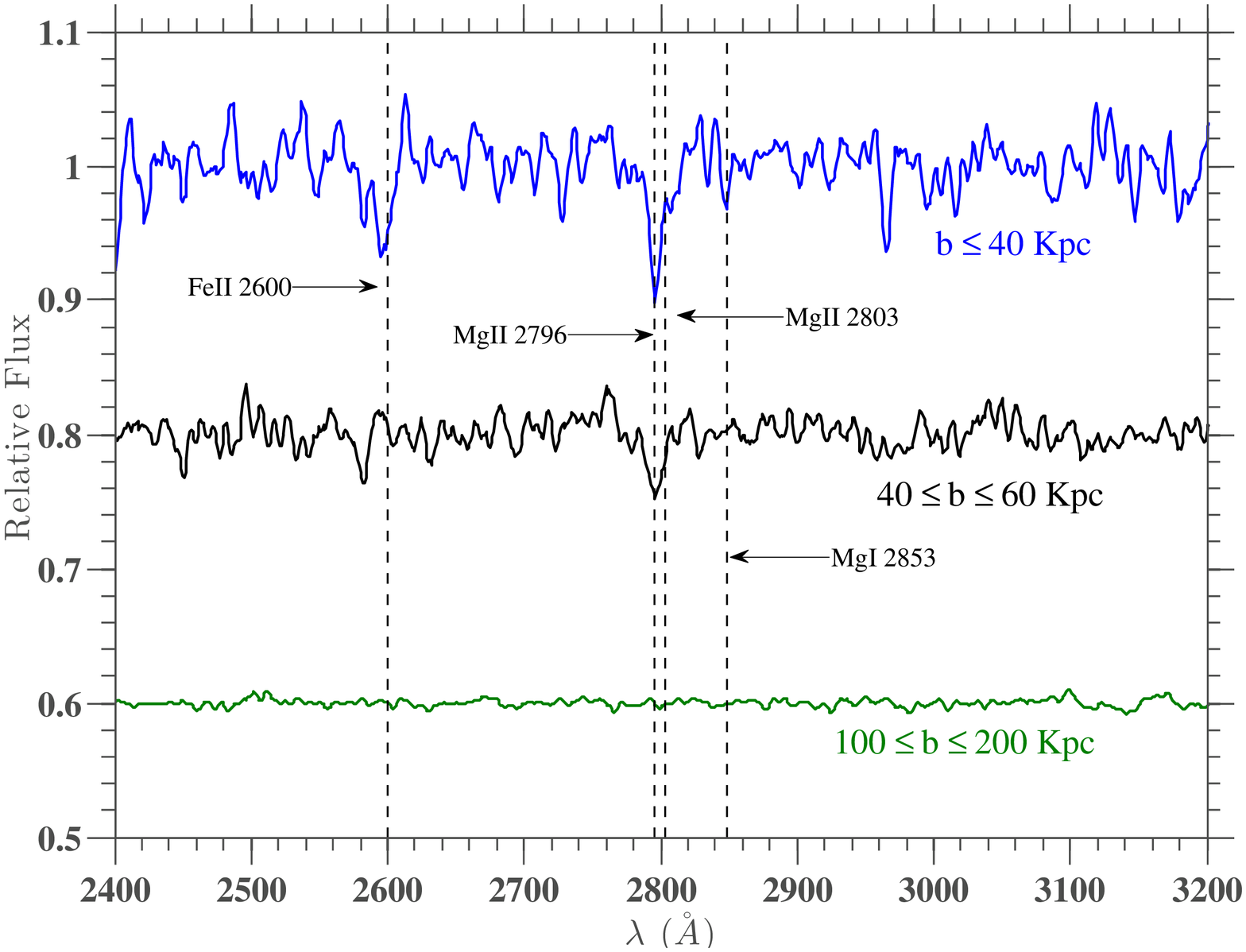}}
\end{center}

\caption{Co-added spectra for all foreground galaxies, at three different impact parameters: $b < 40 $ kpc, $40 \leq b \leq 60$ kpc and $100 \leq b \leq 200$ kpc. The composite spectra are here smoothed with a box of 5 {\AA}  and are displaced vertically for presentation.}
\label{All_coadded_spectra}
\end{figure*}																																						

Clearly a major problem will occur if a ``background'' galaxy actually lies at the same redshift as the foreground galaxy, since then strong interstellar absorption in the former will be mistaken as absorption from the IGM around the latter.  In the current study, this should be unlikely: most galaxies in the background sample lie at $z_{photo} > 1.4$, reflecting the use of the well-established $gzK$ and $ugr$ color-selection criteria in zCOSMOS-deep.  Random scattering to redshifts  $0.5< z < 0.9$ is unlikely.  Catastrophic failures in the photo-z will generally involve degeneracies between $z > 2$ and $z < 0.5$.  The loss of high redshift galaxies with spuriously low photometric redshifts will be of no consequence except to reduce the size of the sample.   The inclusion of low redshift galaxies that are assigned a spuriously high photo-z will merely dilute the signal from the foreground objects, since these sightlines do not in fact penetrate through the foreground system, but will not introduce a false signal.  Using zCOSMOS-deep objects having secured spectroscopic redshifts (i.e. zCOSMOS-deep galaxies with spectroscopic redshift Confidence Classes 4.x and 3.x), we estimate that the fraction of objects in this latter category to be about 3\%. We have not removed these known interlopers, but have checked that none of them lie at the same redshift as the foreground galaxy.
 
We associate to each foreground galaxy a stellar mass and corresponding absolute magnitude as described in the previous section. We divide our foreground galaxies into blue star forming and red passive galaxies on the basis of their rest frame (u-B) color, which is a weak function of mass. The line dividing the blue and red galaxies is similar but not identical to that used in \citep{Peng2010}:
\begin{equation}
\rm{(u-B)_{AB}= 0.98 + 0.075\; log	\left (\frac{M}{10^{10} M_{\odot}} \right ) -0.18z}
\label{mass_color_division}
\end{equation}
where $M$ is the mass of the galaxy in question, for simplicity we use the average redshift $\rm{z \sim 0.7}$ of the sample as a whole.  The color-mass division between blue and red galaxies is shown in Figure \ref{galaxy_color_mass_diagram}. The left hand panel shows this division for non-group galaxies and the right hand panel shows for the group galaxies.  

For environmental information, we use the group catalogue of Knobel et al, in prep. to separate the foreground sample into group and non-group galaxies. Almost 25\% of the initial sample of 7110 galaxies can be assigned to a group. We refer the reader to \cite{Knobel_2009} for a detailed description of the group finding algorithm and details of the zCOSMOS group catalogue.  

To study the azimuthal dependence on Mg II absorption line strength, we use the ZEST morphological classification \citep{Scarlata2007} for each potential foreground galaxy. This is based on the HST/ACS F814W images of the COSMOS field.  We first isolate those foreground galaxies that have a disk-dominated morphology, i.e. ZEST Type 2 excluding bulge-dominated systems, and further require that the galaxy has $0 < b/a < 0.65$, corresponding to inclination angles of $50 < i < 90$. This yields a sample of 595 foreground systems probed by 1134 background spectra within $b < 200$ kpc.

\subsection{Co-addition of galaxy spectra}
The spectra of all background galaxies whose sightlines pass within a given region of a given set of foreground galaxies are then co-added as follows. The spectrum of each individual background galaxy is first shifted in wavelength to the rest-frame of the foreground galaxy in question. This means that absorption at the redshift of the foreground galaxy will appear at the correct "rest-wavelength'', i.e. 2799 {\AA} in the shifted spectrum and will thereby allow co-addition of the spectra of background galaxies associated with foreground galaxies at different wavelengths. The continuum is then normalized by fitting the continuum with a running median filter of width 59 {\AA} and then dividing the spectrum by it. To facilitate co-adding the spectra of different galaxies, this normalized spectrum is  re-sampled onto a uniform grid in wavelength. The co-added spectra is produced by taking the median of the normalized re-sampled spectra. The median is chosen to reduce the sensitivity to absorption or emission features in the background galaxies, and to any other artefacts such as sky residuals, even though the effects of these should be random because of the range of redshifts of both foreground and background sources.  We checked that other summation techniques, including straight averaging and average-sigma-clipping, produce very similar co-added spectra, but marginally less clean continuum. The use of the median, plus the extended nature of the background source means that the absorption in the co-added spectra represents a ``typical" sightline rather than a true average including rare high absorption sightlines.  Three examples of the resulting co-added spectra are shown in Figure \ref{All_coadded_spectra}. 

The spectral resolution of the zCOSMOS-deep spectra is very low ( R $\sim$ 200), so the two components of the Mg II $\lambda \lambda$ 2796, 2803 doublet cannot be separated. The wavelength FWHM of almost 30 {\AA} is much larger than the observed-frame 12 {\AA} separation of the doublet at the redshifts of interest. We therefore fit a single Gaussian profile around the Mg II rest-frame wavelength and measure the (rest-frame) absorption equivalent width, integrating across both components of the doublet.  In order to limit the number of free parameters in the fit, we determine the shape of this Gaussian (central wavelength and dispersion) from a fit to the best absorption spectrum, which is obtained within 40 kpc of all foreground galaxies (shown in Figure \ref{All_coadded_spectra}, top spectra).  This Gaussian profile is then used for all other spectra, with only the depth as a free parameter. We have checked that other integration schemes, e.g. simply summing the flux deficit through the region of the line, give indistinguishable results within the uncertainties.

The errors on the equivalent width measurements are determined using a bootstrap approach.  For each set of background spectra, a thousand co-added spectra are  generated from random selections of the sample. The width of the distribution of measured equivalent widths in these thousand spectra are taken as representative of the error in the original measurement, and should account for sample variance, continuum uncertainties and profile fitting errors.

\begin{figure*}[htb]
\begin{center}
{\includegraphics[angle=0,width=16cm,height=11cm]{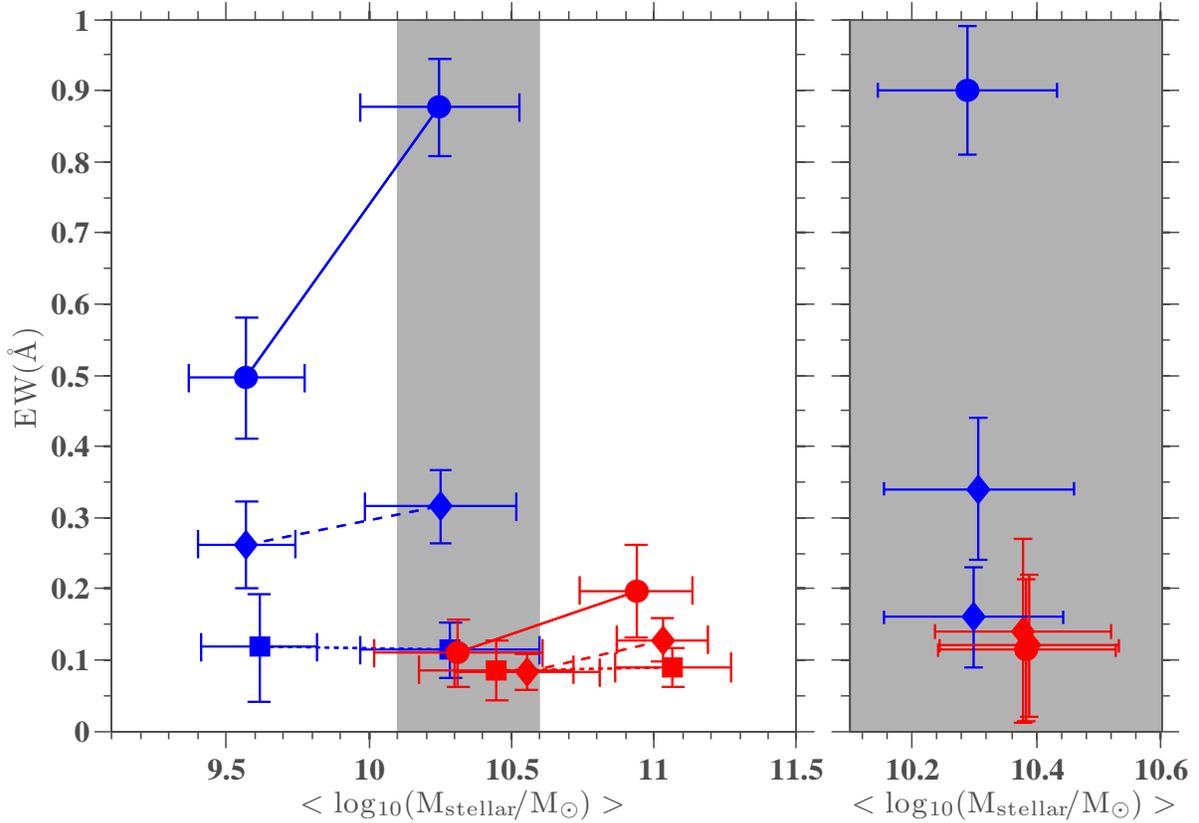}}

\end{center}

\caption{Comparison of Mg II EW for different samples of the foreground galaxies selected by their mass and rest frame color. The blue and red points are for blue and red galaxies respectively. Three impact parameter bins in the range of 0-50 kpc (filled circles),  50-65 kpc (diamonds) and 65-80 kpc ( squares) are used. Evidently there is a strong dependence on color for low impact parameter bins.  This effect is not visible as we probe further away from the galaxy. There is also a mass dependence on the blue sample which is not evident for the red sample. In the left panel, we apply a small offset of 0.1 in mass for clarity. The shaded region in both panels indicate the common mass bin chosen to compare the color dependence on Mg II absorption strength.}
\label{four_mass_color_sample}
\end{figure*}					


\begin{deluxetable*}{l c c c c c}
\tablewidth{0pt}
\tabletypesize{\footnotesize}
\tablecaption{Measured equivalent width of the complete sample as a function of stellar mass and rest frame color.}
\tablehead{
\colhead{Sample} &
\colhead{b\tablenotemark{a}} &
\colhead{$\rm{<mass>}$\tablenotemark{b}} &
\colhead{Number\tablenotemark{c}}  &
\colhead{EW\tablenotemark{d}} &
}
\startdata
 &$b < 50$ & 9.57 $\pm$ 0.20  & 116 & 0.50 $\pm$ 0.08   \\
Blue galaxies ($M_{stellar} < 9.88\; M_{\odot}$) & $50 < b < 65$ & 9.57 $\pm$ 0.17 & 90 & 0.26 $\pm$ 0.06    \\ 
 & $65 < b < 80$& 9.6 $\pm$ 0.20 &  126 & 0.12 $\pm$ 0.08  \\
 \\
 \hline
\\
  & $b < 50$& 10.25 $\pm$ 0.28 &  141 & 0.88 $\pm$ 0.07\\
Blue galaxies ($M_{stellar} \geq 9.88\;M_{\odot}$)& $50 < b < 65$ & 10.25 $\pm$ 0.27
 &  122 & 0.32 $\pm$ 0.05    \\
 & $65 < b < 80$ & 10.28 $\pm$ 0.32  & 154 & 0.11 $\pm$ 0.04   \\
\\
\hline 
\\
 & $b < 50$ & 10.31$\pm$ 0.30 & 104 & 0.11 $\pm$ 0.05   \\
Red galaxies ($M_{stellar} < 10.68\;M_{\odot}$) & $50 < b < 65$ & 10.35 $\pm$ 0.25 & 75 & 0.08 $\pm$ 0.03   \\
 & $65 < b < 80$ & 10.35 $\pm$ 0.27
& 102 & 0.09 $\pm$ 0.04   \\
\\
\hline
\\
  & $b < 50$ & 10.94 $\pm$ 0.20  & 99 & 0.20 $\pm$ 0.07  \\
Red galaxies ($M_{stellar} \geq 10.68\;M_{\odot}$) & $50 < b < 65$ & 10.93 $\pm$ 0.16  & 84 & 0.13 $\pm$ 0.03  \\
 &$65 < b < 80$ & 10.97 $\pm$ 0.21 & 115 & 0.09 $\pm$ 0.03   \\
\enddata
\tablenotetext{a}{Range of impact parameter, in physical kpc.}
\tablenotetext{b}{Mean mass of the foreground galaxies co-added, in $\rm{M_{stellar}/M_{\odot}}$. }
\tablenotetext{c}{Measured rest frame equivalent width in \AA. }
\tablenotetext{d}{Number of foreground-background pairs.}
\label{table:sample_table1}
\end{deluxetable*}				
				

\section{Results}

In the following sections we discuss the variation of Mg II absorption equivalent width with different properties of the foreground galaxies, examining the radial profile ,for the inclined disk galaxies, the azimuthal dependence and for disk galaxies, the dependence on apparent inclination.
																																										
\subsection{Radial dependence and variation with stellar mass and color}

We first investigate the variation of Mg II line strength with impact parameter as a function of the color and mass of the foreground galaxy. The foreground sample is divided into four sub-samples in terms of their stellar masses and rest frame colors: A high-mass blue sample is defined with $\rm{\log_{10}(M_{stellar}) \geq 9.88\; M_{\odot}}$, a low mass blue sample comprises galaxies with $\rm{\log_{10}(M_{stellar}) < 9.88 \;M_{\odot}}$. A high mass red sample is defined with $\rm{\log_{10}(M_{stellar}) \geq 10.68\; M_{\odot}}$ and a low mass red sample comprises galaxies with $\rm{\log_{10}(M_{stellar}) < 10.68\; M_{\odot}}$, as shown in Figure \ref{galaxy_color_mass_diagram}. These mass ranges are chosen so that each sample has approximately the same number of background spectra and so that the high mass blue sample and the low mass red sample have more or less the same mean stellar mass, thereby enabling a direct comparison between blue and red galaxies at the same mass.  For each sample, a co-added spectrum is produced for impact parameters $b < 50$ kpc, $50 < b < 65$ kpc and $65 < b < 80$ kpc.

The absorption equivalent widths measured in the twelve co-added spectra are plotted in Figure \ref{four_mass_color_sample} and displayed in Table \ref{table:sample_table1}. The error bars on mass give the $1 \sigma$ spread of the foreground galaxy mass in that bin. It can clearly be seen that blue foreground galaxies are associated with stronger Mg II absorption relative to red galaxies, especially at small impact parameters. To study this in more detail, we also create a single relatively narrow intermediate bin centered in mass on the overlap region, i.e. ($\rm{10.1 \;M_{\odot} \geq \log_{10}(M_{stellar}) \geq 10.6 \;M_{\odot}}$, shaded region Figure \ref{four_mass_color_sample}). This is shown in the panel to the right.  In this overlap region of mass, the blue galaxies have almost eight times larger absorption at $b < 50$ kpc. There is also a significant mass dependence within the blue sample, again most clearly seen at smaller impact parameters. This may be present for the red sample but the statistical significance is very low. 

These results are consistent with the findings of \cite{2007Zibetti}, who used the orthogonal approach of stacking {\it images} from SDSS to show that stronger absorption systems in quasar spectra were associated with bluer, star-forming, galaxies.   The current analysis provides a direct proof of the dependence of Mg II absorption on the colors of the associated galaxies.   

This clear demonstration of a color dependence is consistent with the scenario that Mg II absorption within 65 kpc of galaxies is associated with material entrained in outflows driven by star-formation. This is also indicated by analysis of blue-shifted Mg II absorption profiles in the spectra of star-forming galaxies \citep{Wiener2009, Rubin2010}. However, this evidence is also consistent with the idea that in-falling gas triggers star formation activity in blue galaxies.  We show later in the paper, however, that the azimuthal dependence of the absorption strongly favors the first hypothesis.

Remembering that our equivalent width measurements are averaged over the annuli surrounding the foreground galaxies, it is noticeable how strong this average absorption is around the most massive blue galaxies, even allowing for the fact that we compute integrated EW across both components of the doublet.  In quasar spectra (which individually probe a much smaller region) an equivalent width of greater than 0.3 {\AA} is usually regarded as a ``strong'' system.  This level of absorption (i.e. integrated EW $>$ 0.6 {\AA} is attained across an area of more than 8000 kpc$^2$ around the blue galaxies with masses around $2 \times 10^{10}$ M$_{\odot}$. 
																							
\begin{figure*}
\begin{center}
{\includegraphics[angle=0,width=8.9cm,height=8.2cm]{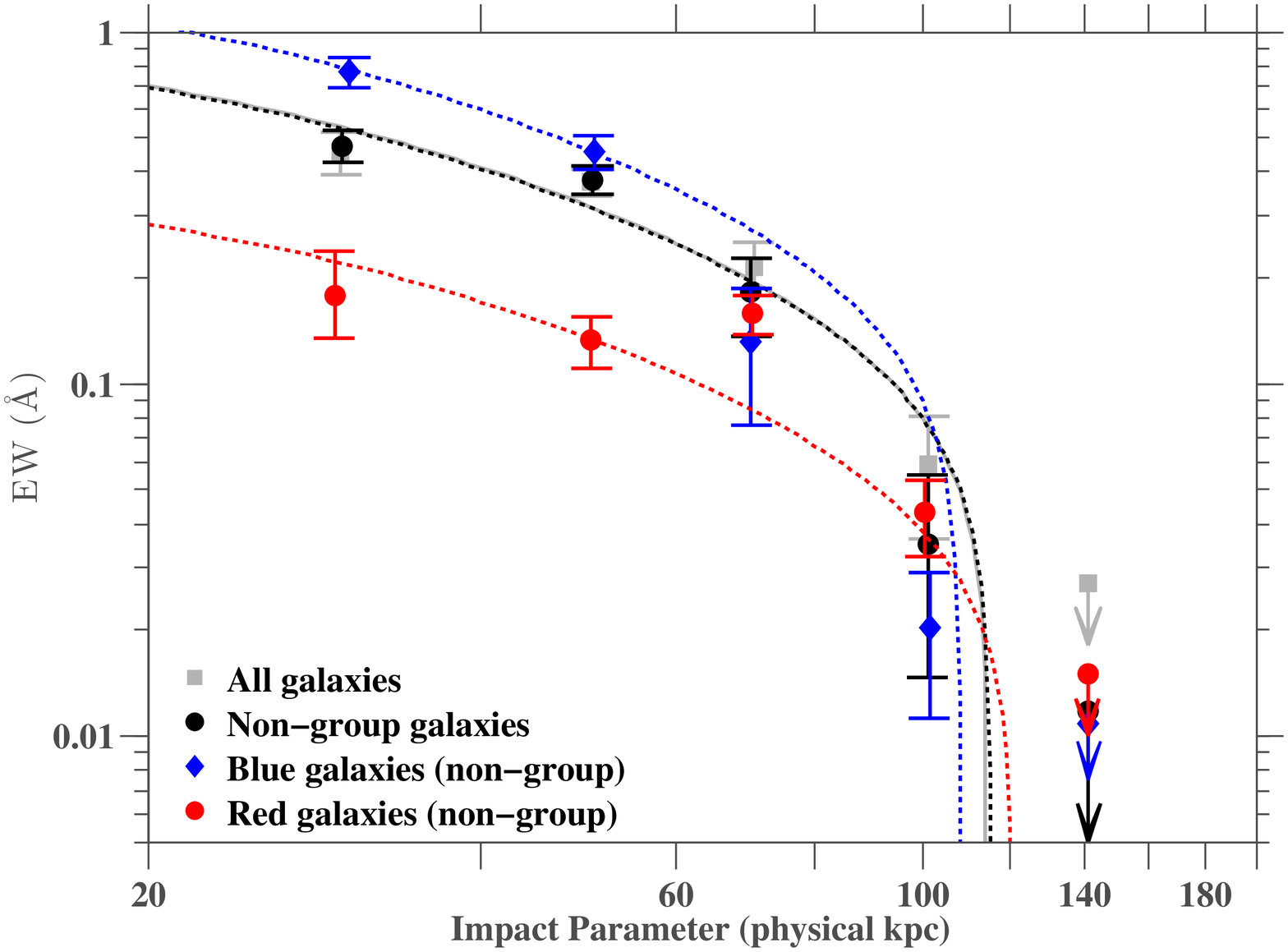}}
{\includegraphics[angle=0,width=8.9cm,height=8.2cm]{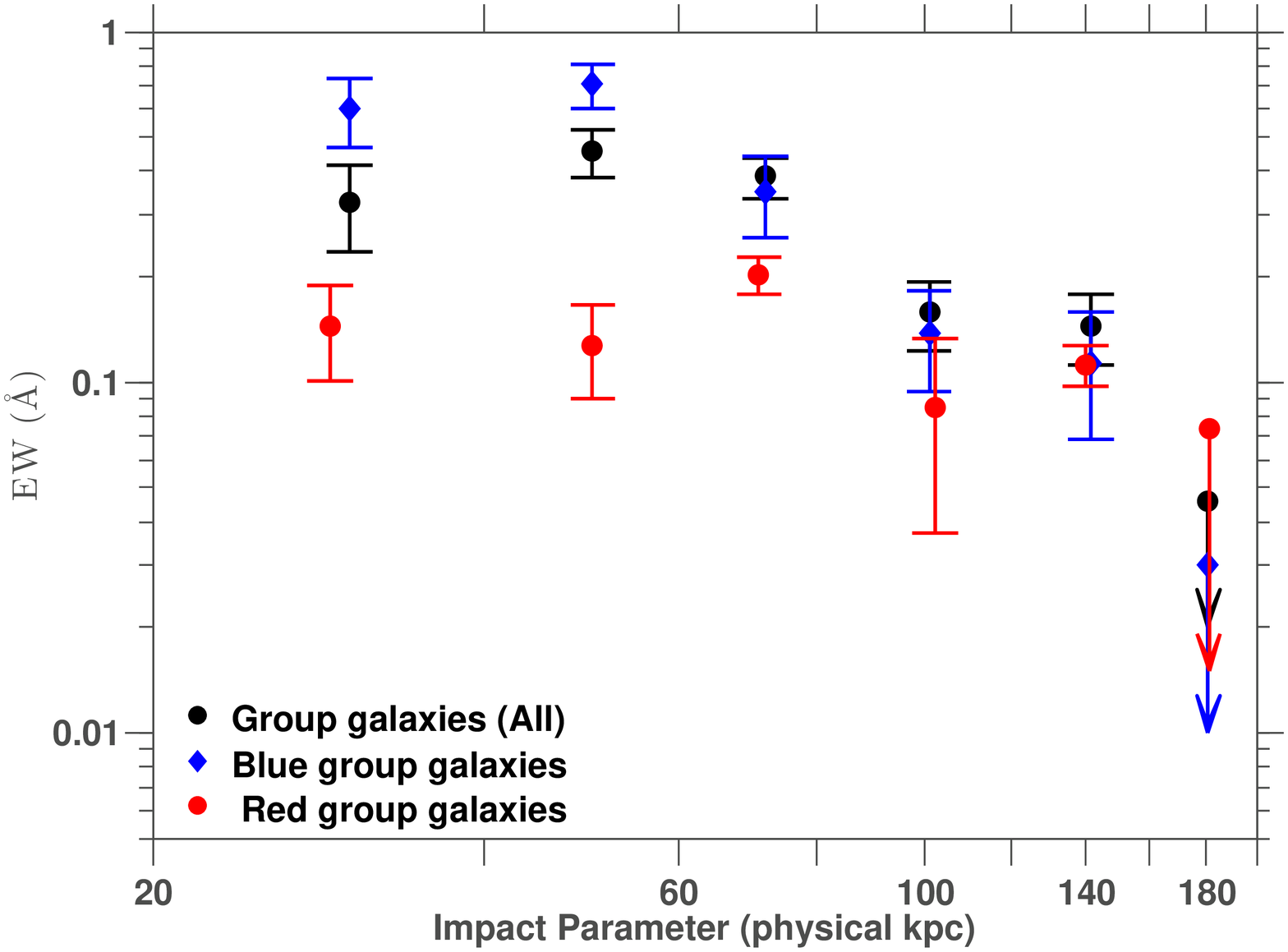}}
\caption{Mg II radial profiles around galaxies, the left panel is for non-group galaxies and the right panel is for group galaxies. The points are measured equivalent widths for the corresponding impact parameters. The dotted lines are fitted SIS profile as described in section 3.2. The red and blue colors correspond to red and blue galaxies and the black color corresponds to all non-group galaxies in the left panel and all group-galaxies in the right panel. The grey square points in the left panel are for all galaxies.}
\label{All_object_EW_field_red_blue_mass}
\end{center}
\end{figure*}				

										
\begin{deluxetable*}{l c c c c c}[h!]
\tablewidth{0pt}
\tabletypesize{\footnotesize}
\tablecaption{Mg II absorption radial profiles around non-group/group galaxies}
\tablehead{
\colhead{Environment\tablenotemark{a}} &
\colhead{Color\tablenotemark{b}} &
\colhead{$\rm{b}$\tablenotemark{c}} &
\colhead{Number\tablenotemark{d}} &
\colhead{EW\tablenotemark{e}}  &
}
\startdata
 & & $0 < b < 40$  & 198 & 0.48 $\pm$ 0.05 \\
 & & $40 < b < 60$ & 300 & 0.38 $\pm$ 0.05\\
 & All galaxies &  $60 < b < 80$  & 475 & 0.18 $\pm$ 0.05 \\
 & &  $80 < b < 120$  & 1273 & 0.04 $\pm$ 0.02\\
 & &  $120 < b < 160$   & 1785 & < 0.012\\
 \\
 \cline{2-5}
 \\
 & &  $0 < b < 40$  & 116 & 0.77 $\pm$ 0.08\\
 & &  $40 < b < 60$   & 159 & 0.46 $\pm$ 0.05\\
Non-group galaxies & Blue galaxies &  $60 < b < 80$   & 278 & 0.13$\pm$ 0.06\\
 & &  $80 < b < 120$   & 742 & 0.02 $\pm$ 0.01\\
 & &  $120 < b < 160$  & 1075 & < 0.011 \\
 \\
\cline{2-5}
 \\
 & &  $0 < b < 40$  & 82 & 0.18 $\pm$ 0.05 \\
 & &  $40 < b < 60$   & 141 & 0.13 $\pm$ 0.02\\
 & Red galaxies&  $60 < b < 80$  & 197  & 0.16 $\pm$ 0.02 \\
 & &  $80 < b < 120$   & 531 & 0.04 $\pm$ 0.01\\
 & &  $120 < b < 160$   & 710 & < 0.015\\
 \\
 \hline
 \\
 & &  $0 < b < 40$   & 75 & 0.33 $\pm$ 0.09 \\
 & &  $40 < b < 60$   & 116 & 0.46 $\pm$ 0.07\\
 & All galaxies &  $60 < b < 80$ & 196  & 0.39 $\pm$ 0.05 \\
 & &  $80 < b < 120$  & 524 &  0.16 $\pm$ 0.04\\
 & &  $120 < b < 160$   & 669 & 0.15 $\pm$ 0.03\\
 & &  $160 < b < 200$  & 944  & < 0.046\\
 \\
 \cline{2-5}
 \\
 & &  $0 < b < 40$   & 36 & 0.61 $\pm$ 0.13\\
 & &  $40 < b < 60$  & 62 & 0.71 $\pm$ 0.10 \\
 Group galaxies & Blue galaxies &  $60 < b < 80$   & 92 & 0.35 $\pm$ 0.09\\
 & &  $80 < b < 120$  & 248 & 0.14 $\pm$ 0.04 \\
 & &  $120 < b < 160$   & 320 & 0.11 $\pm$ 0.05\\
 & &  $160 < b < 200$  & 467 & < 0.03\\
 \\
 \cline{2-5}
 \\
 & &  $0 < b < 40$   & 39 & 0.15 $\pm$ 0.04\\
 & &  $40 < b < 60$  & 54 & 0.13 $\pm$ 0.04 \\
 & Red galaxies&  $60 < b < 80$  & 104  & 0.20 $\pm$ 0.02\\
 & &  $80 < b < 120$  & 276 & 0.09 $\pm$ 0.05 \\
 & &  $120 < b < 160$  & 349 & 0.11 $\pm$ 0.02 \\
 & &  $160 < b < 200$  & 477 & < 0.074 \\
\tablenotetext{a}{Dividing the galaxies in terms of environment.}
\tablenotetext{b}{Galaxies are divided as blue and red, in terms of their rest frame colors.}
\tablenotetext{c}{Range of impact parameter, in physical kpc.}
\tablenotetext{d}{Number of foreground-background pairs.}
\tablenotetext{e}{Measured rest frame equivalent width in \AA. }
\label{table:sample_table2}
\end{deluxetable*}				

\subsection{The radial dependence in group and non-group environments}

To study further the radial profile around galaxies, we divide the sample more finely in $b$ but lump together all blue galaxies and all red galaxies.  We also distinguish between galaxies that are identified as lying in groups (as described in Section 2) and those not in groups. The resulting radial profiles are shown in the Figure \ref{All_object_EW_field_red_blue_mass} and also displayed in Table \ref{table:sample_table2}.

The non-group galaxies, in the left hand panel, show a profile that drops steeply with radius beyond about 70 kpc. To ease comparison with previous work, we  parametrize the radial profiles in terms of a singular isothermal sphere (SIS) as in \cite{2008T&C}.  We show later in Section 3.3 that the Mg II absorption is certainly not spherically distributed, and thus our choice has no physical meaning per se, and is used simply as a convenient parametrization. In a  future paper we will explore more physically motivated models for the radial profile.

In the isothermal sphere distribution, the distribution of Mg II gas is characterized as
				
\begin{equation}
\rm{EW_{r}^{iso}  =
\begin{cases}
  \frac{EW_{0}}{\sqrt{(b^{2}/a_{h}^{2} + 1)}} \arctan {\sqrt{\frac{R_{gas}^{2} - b^{2}}{b^{2} + a_{h}^{2}}}} , & \mbox{if }b \leq R_{gas} \\ \\
 0, & \mbox{otherwise }
\end{cases}
}
\end{equation} 	

Here the core radius $\rm{a_{h}}$ is defined to be $\rm{a_{h} = 0.2 R_{gas}}$ and does not affect $\rm{EW_{r}^{iso}}$ at large b. We fit this model to the equivalent width profiles of the non-group galaxies. We find
$\rm{R_{gas}}$ = 115.2 $\pm$ 2.3 kpc (physical) and $\rm{EW_{0}}$ = 0.7 $\pm$ 0.12 for ``all'' galaxies, $\rm{R_{gas}}$ = 107.6 $\pm$ 1.3 kpc (physical) and $\rm{EW_{0}}$ = 1.1 $\pm$ 0.14 for blue galaxies and $\rm{R_{gas}}$ = 118.1 $\pm$ 5. kpc (physical) and $\rm{EW_{0}}$ = 0.3 $\pm$ 0.11 for red galaxies.  
For completeness, the results for all galaxies (group plus non-group) is very similar $\rm{R_{gas}}$ = 114.7 $\pm$ 1.5 kpc (physical) and $\rm{EW_{0}}$ = 0.71 $\pm$ 0.12. It is noticeable that the $R_{gas}$ for the blue and the red galaxies are very similar despite large changes in $EW_{0}$.
					
Although the methodologies are different, in the sense that we measure a spatially averaged absorption, our own analysis clearly confirms the trend between between impact parameter (b) and equivalent width (EW) for galaxies, for example as observed in \cite{Chen2010a} with quasar absorption lines in $z < 0.5$ SDSS galaxies.  However \cite{Chen2010a} did not find any color dependence for non-group $z < 0.5$ galaxies. As shown in Figure \ref{four_mass_color_sample}, our co-added spectra (at $<z> \sim 0.7$) show a strong dependence on the color of the host galaxy for $\rm{b \leq 70\; kpc}$. The reason for this discrepancy is unclear to us.  It might reflect the higher redshifts, and thus stronger star-formation activity, of blue galaxies in our own sample, but the redshift difference is small enough to make this interpretation unlikely in our view.

The right hand panel in Figure \ref{All_object_EW_field_red_blue_mass} shows the radial profile around galaxies, which are selected to lie in groups. This is somewhat artificial, essentially because the absorption from each line of sight will contribute to several different $b$ bins, corresponding to the different members of the group. This will automatically produce a flatter distribution of Mg II with impact parameter.  We include it in this paper only for comparison with previous work (e.g. \citealt{Chen2010a}). With this caveat in mind,
we see a much flatter absorption profile that extends beyond 140 kpc, substantially larger than seen for isolated, non-group galaxies.  It should be noted that the dependence on color that we have observed for non-group galaxies is also evident in group galaxies at small radii $b < 70$ kpc but is much weaker at larger radii.  

\begin{figure*}
\begin{center}
{\includegraphics[angle=0,width=8.9cm,height=8.2cm]{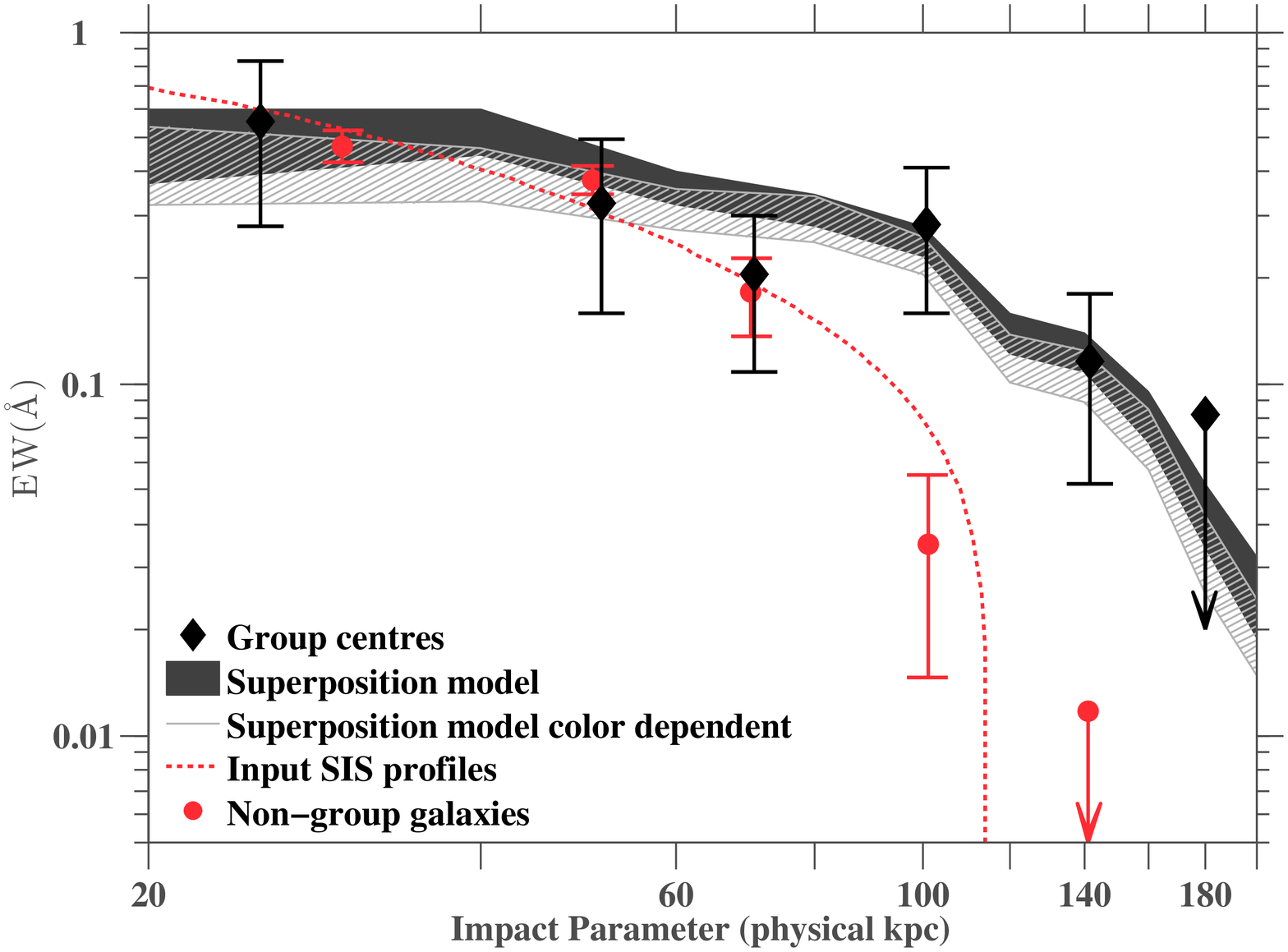}}
{\includegraphics[angle=0,width=8.9cm,height=8.2cm]{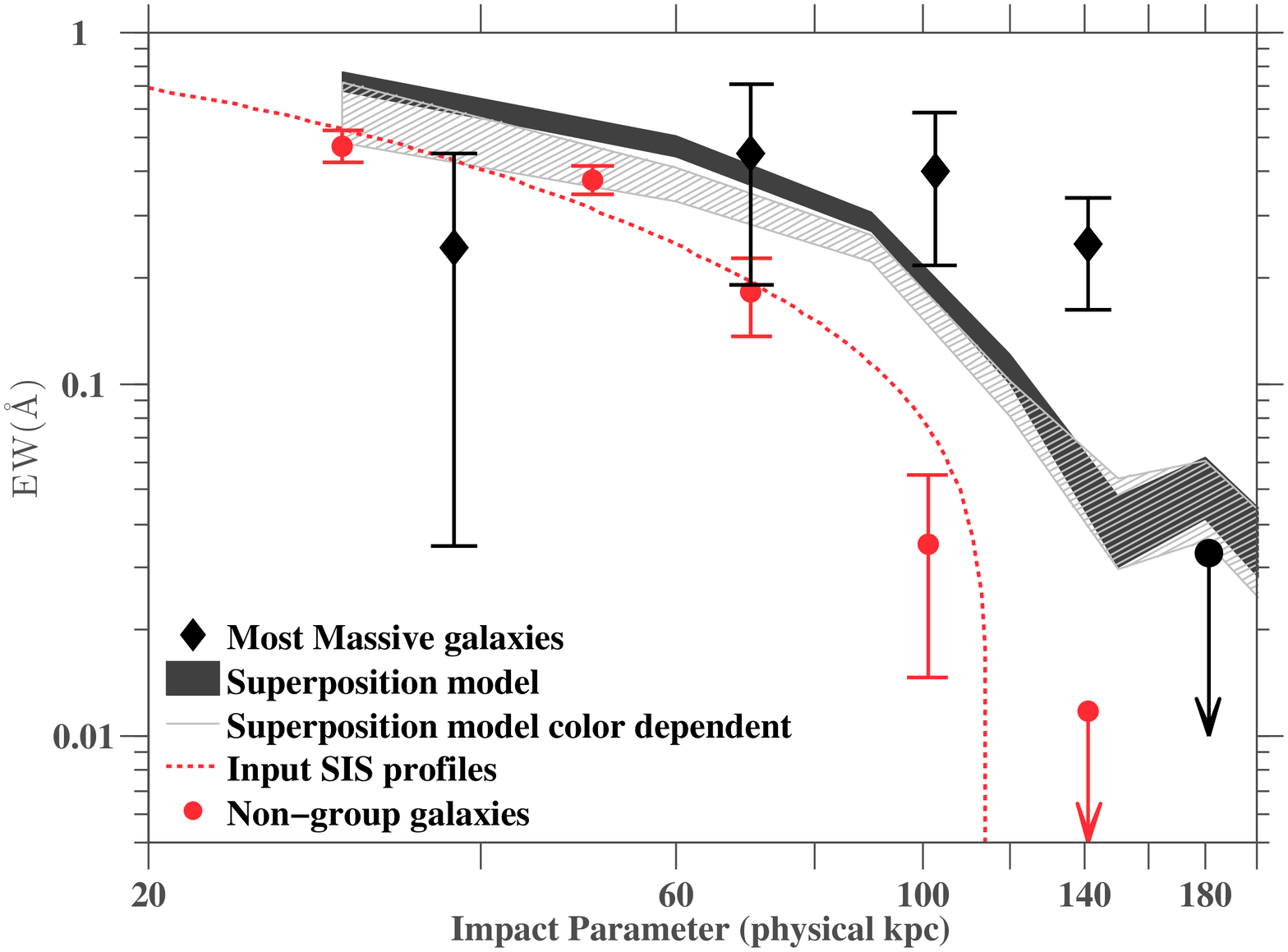}}

\caption{The average Mg II radial profile around the geometric group-centres (left panel) and around the most massive galaxy in the groups (right panel). The dotted line in both panels is the average fitted radial profile for non-group galaxies and the filled circles are measured radial profile of isolated galaxies. The black diamonds in the left panel give the measured radial profile of group centres and the black diamonds in the right panel give the measured radial profile for most massive galaxies in the groups. In both panels the dark grey filled region and the light grey filled region are the expected radial profiles from the superposition model if we assign each member galaxy an average Mg II absorption profile and if we assign different Mg II absorption profiles to each member galaxy depending on their rest frame color respectively. }
\label{Group_center_toy_model_prediction}
\end{center}
\end{figure*}

A more physically motivated approach would be to determine the Mg II absorption profile around a particular location in the group, e.g. either the geometric center defined in Knobel et al in prep. as the geometric mean of the positions in the sky of all the group members, or the location of the most massive member of the group. The results are shown in Figure \ref{Group_center_toy_model_prediction}. The left panel is for group centres and the right panel is for the most massive galaxies in the groups.

\begin{figure*}
\begin{center}
{\includegraphics[angle=0,width=12cm,height=10cm]{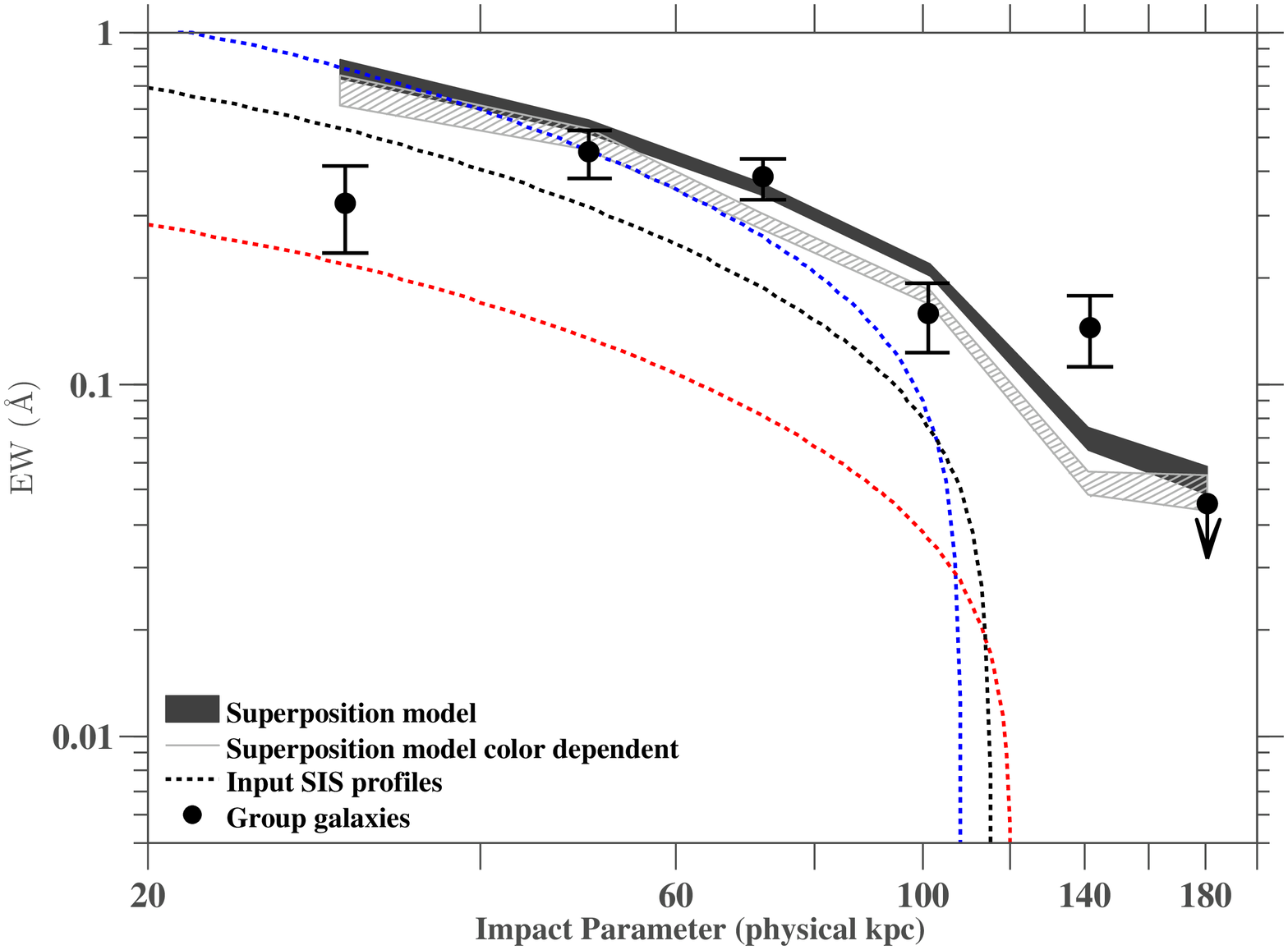}}
\end{center}

\caption{The black points with filled circles are EW measurements for all group galaxies, the dotted lines are the input SIS profiles for all galaxies (black dotted line), blue galaxies (blue dotted line) and red galaxies (red dotted line) respectively. The dark shaded region gives the EW profile expected from the superposition model with a radial SIS profile for each halo as given by the black dotted line. The light shaded region is the expectation of the same model if we include the color information for individual group members and assign different EW profiles for red and blue galaxies as given by the dotted red and blue lines.}
\label{Group_toy_model_prediction}
\end{figure*}																																						

In order to try to understand the profiles in the right hand panel of Figure \ref{All_object_EW_field_red_blue_mass} and in  Figure \ref{Group_center_toy_model_prediction}, we construct a simple model in which the absorption profile around each group galaxy is given by that which we have measured around non-group galaxies - i.e. the model assumes that the absorption of the group can be represented by a simple superposition of the absorption of individual members. Knowing the locations of each individual members within each group, we can predict, for the ensemble of group members, the average profile that would be expected if all group members exhibited the same average profile as non-group members, as described by the SIS profiles in the left hand panel of Figure \ref{All_object_EW_field_red_blue_mass}. On average only 2/3 of the $I_{AB} \leq 22.5$ galaxies were observed spectroscopically in zCOSMOS-bright. Knobel et al in prep. have developed a scheme to include in the groups, the missing members for which only a photo-z is available. In the model we include along with the spectroscopic members, the photometric members with a probability of membership $p >0.7$. We first ignore the color of the other group members, but then incorporate this information also into the model.

If, within some bin in impact parameter, we have a sample of $n$ group members, which are spread amongst some number of groups, and if the $n$th galaxy in this sample has $m$ fellow members of its particular group, then the final equivalent width that is expected for this impact parameter is given by:
\begin{equation}
\rm{EW_{total} = \frac{1}{n} \sum_{n} ( \sum_{m} EW_m(b_m))}
\end{equation}			
																																	
where $EW_m(b_m)$ is the equivalent width appropriate for the SIS sphere of the galaxy type for galaxy $m$ (i.e. blue or red) at the impact parameter $b_m$ that the $background$ galaxy has to this $m$th member galaxy. In summing the equivalent widths, we assume that the velocities associated with each galaxy are large enough that there are no saturation effects.  Typical velocity dispersions within the groups in question are typically $200 < \sigma < 600$ kms$^{-1}$ \citep{Knobel_2009}.

The results of this exercise are shown in Figure \ref{Group_toy_model_prediction} and in Figure \ref{Group_center_toy_model_prediction}. Figure \ref{Group_toy_model_prediction}  re-plots the average group member profile of Figure \ref{All_object_EW_field_red_blue_mass} but no longer differentiating in color of the foreground galaxies. Both in Figure \ref{Group_toy_model_prediction} and in Figure \ref{Group_center_toy_model_prediction}, the dotted lines show what would be expected from the SIS model around a single galaxy, i.e. ignoring other group members entirely, the dark shaded area include the effects of other group members using the superposition approach, while the light shaded area also incorporate the information on the colors of those other members.  Both of the latter shaded areas provide a better representation of the extended profile observed in group galaxies.   This simple superposition model provides a reasonable representation of the absorption profiles around group galaxies.  Interestingly, such a model would predict that the absorption profile at $b > 100$ kpc should be independent of the color of the member galaxy, since at these radii it would be dominated by other group members.  This may indeed be seen in the right hand panel of Figure \ref{All_object_EW_field_red_blue_mass}.

This superposition model reproduces both the extended distribution and  the magnitude of the Mg II absorption profile very well using the geometrical group centers. However, while using the most massive galaxies, the result is less satisfactory.  The differences between the two panels of Figure \ref{Group_center_toy_model_prediction} arise from differences in both the models and in the observational data points, especially at large impact parameters.    The difference in the models arises because the group centers are, by construction, more or less equidistant from the group members, maximizing the number which are close enough to contribute to the expected EW.  Within the data, there are about 2.5 times more group members within 100 kpc of the geometrical center as there are within 100 kpc of the most massive galaxies.  The differences in the observational points are not of large significance compared with the statistical error bars.  The offsets between the geometrical centres and the most massive galaxies in each group extend up to 140 kpc, producing substantial scrambling of the data in impact parameter between the two approaches.  Therefore the error-bars between the plots should be largely independent.

The success of this superposition model, especially for geometrical group centres is quite surprising as it suggests that the existence of Mg II absorption haloes around galaxies may not be significantly affected by the group environment.  The fact that the specific star-formation rates of star-forming galaxies do not appear to depend on environment, even though the fraction of galaxies that are star-forming does \citep{Peng2010}, suggests that the source of star-formation driven winds is likely the same in the two environments. However, tidal effects have been invoked to enhance absorption for low redshift groups \citep{Kacprazak2010b}. One can also imagine scenarios whereby different mechanisms may counteract each other, and the accuracy of the agreement anyway allows for some non-negligible differences.

\begin{deluxetable}{l c c c c }
\tablewidth{0pt}
\tabletypesize{\footnotesize}
\tablecaption{Radial profile of Mg II absorption around disk galaxies at different azimuthal angles.}
\tablehead{
\colhead{$\rm{\phi}$\tablenotemark{a}} &
\colhead{b\tablenotemark{b}} &
\colhead{Number\tablenotemark{c}}  &
\colhead{EW\tablenotemark{d}}  &
}
\startdata
& $0 < b < 35$ & 23 & 1.17 $\pm$ 0.25\\
$0\,^{\circ} < | \phi| < 45\,^{\circ}$& $35 < b < 50$  & 31 & 0.92 $\pm$ 0.25 \\
&  $50 < b < 65$ & 42 &  0.47 $\pm$ 0.12\\
&  $65 < b < 80$ & 91 & 0.19 $\pm$ 0.10 \\
\\
\hline
\\
& $0 < b < 35$ & 25 & 0.41 $\pm$ 0.20 \\
$45\,^{\circ} < | \phi| < 90\,^{\circ}$ & $35 < b < 50$  & 28 &  0.47 $\pm$ 0.14\\
&  $50 < b < 65$ & 49 & 0.43 $\pm$ 0.10\\
&  $65 < b < 80$& 93 & 0.24 $\pm$ 0.10\\
\enddata
\tablenotetext{b}{Range of azimuthal angles considered, in degrees}
\tablenotetext{a}{Range of impact parameter, in physical kpc.}
\tablenotetext{c}{Number of foreground background pairs. }
\tablenotetext{d}{Measured rest frame equivalent width in \AA. }
\label{table:sample_table3}
\end{deluxetable}

\begin{deluxetable}{l c c c c }
\tablewidth{0pt}
\tabletypesize{\footnotesize}
\tablecaption{Azimuthal profile of Mg II absorption around disk galaxies at different impact parameters.}
\tablehead{
\colhead{b\tablenotemark{a}} &
\colhead{$\rm{\phi}$\tablenotemark{b}} &
\colhead{Number\tablenotemark{c}}  &
\colhead{EW\tablenotemark{d}}  &
}
\startdata
& $0\,^{\circ} < |\phi| < 30\,^{\circ}$ & 25 & 1.04 $\pm$ 0.15 \\
 b $<$ 40 kpc & $30\,^{\circ} < |\phi| < 60\,^{\circ}$ & 26 & 0.77 $\pm$ 0.14\\
& $60\,^{\circ} < |\phi| < 90\,^{\circ}$ & 31 & 0.32 $\pm$ 0.19\\
\\
\hline
\\
& $0\,^{\circ} < |\phi| < 30\,^{\circ}$ & 32 & 0.60 $\pm$ 0.17  \\
40 $<$ b < 60 kpc & $30\,^{\circ} < |\phi| < 60\,^{\circ}$ & 39 & 0.64 $\pm$ 0.14\\
& $60\,^{\circ} < |\phi| < 90\,^{\circ}$ & 38 & 0.45 $\pm$ 0.13\\
\\
\hline
\\
& $0\,^{\circ} < |\phi| < 30\,^{\circ}$ & 60 &  0.19 $\pm$ 0.12\\
60 $<$ b < 80 kpc & $30\,^{\circ} < |\phi| < 60\,^{\circ}$ & 65& 0.36 $\pm$ 0.11 \\
& $60\,^{\circ} < |\phi| < 90\,^{\circ}$ & 66 & 0.28 $\pm$ 0.15\\
\enddata
\tablenotetext{a}{Range of impact parameter, in physical kpc.}
\tablenotetext{b}{Range of azimuthal angles considered, in degrees}
\tablenotetext{c}{Number of foreground background pairs. }
\tablenotetext{d}{Measured rest frame equivalent width in \AA. }
\label{table:sample_table4}
\end{deluxetable}

\begin{figure*}
\begin{center}
{\includegraphics[angle=0,width=12cm,height=9.6cm]{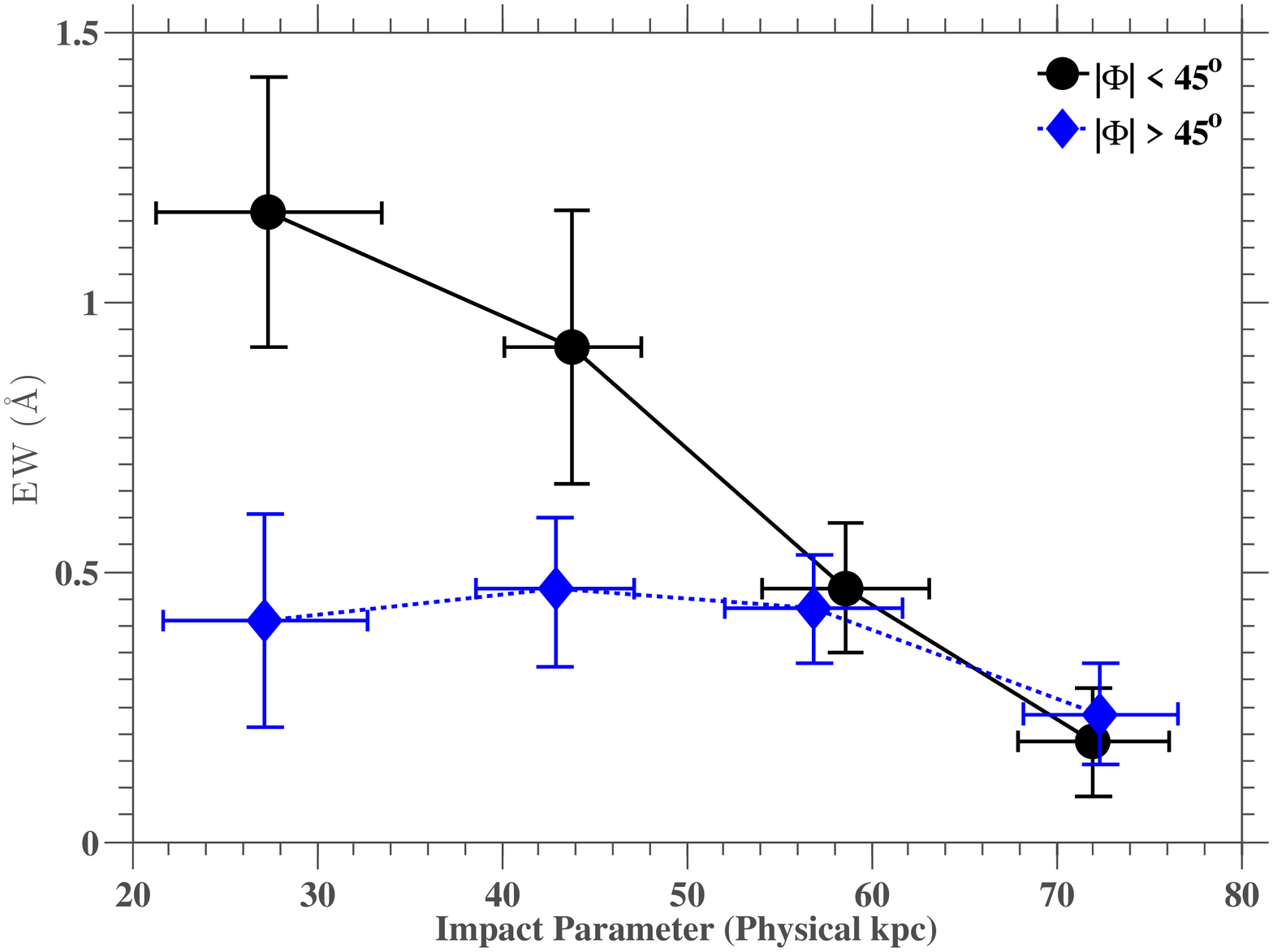}}
\caption{\label{Bipolar_outflow_plot}Dependence of Mg II EW on the azimuthal angle relative to the disk axis. The filled black circles with solid line are the measured Mg II equivalent width as a function of impact parameter around disk galaxies close to the axis of the disk (i.e. $| \phi | < 45\,^{\circ}$) and the dashed line with filled diamonds give the measured Mg II equivalent width near the plane of the disk (i.e. $| \phi | > 45\,^{\circ}$) . For low impact parameters, stronger Mg II systems are observed close to the axis of the disk as compared to the plane of the disk. This effect diminishes as higher impact parameters are probed and is no longer distinguishable for very high impact parameters. The errorbars in impact parameter give the standard deviation of the distribution of impact parameters within that bin.}
\end{center}
\end{figure*}			
						
\subsection {The azimuthal absorption profile around disk galaxies}																											
With the current dataset we can probe the strength of Mg II absorption lines around disk galaxies as a function of the azimuthal angle relative to the projected disk axis. As described in Section 2, we select a set of disk-dominated galaxies that lie within $40\,^{\circ} $ of being edge on.  We then compute the azimuthal angle $\phi$ between the projected semi-minor axis of the disk, and the projected vector from the center of the foreground galaxy to the background galaxy.  In other words, values of $|\phi| > 45\,^{\circ} $ represent lines of sight that pass near to the plane of the disk, while values of of $|\phi| < 45\,^{\circ}$ are associated with lines of sight passing close to the symmetry (rotation) axis of the disk.  
If the Mg II absorption around disk galaxies is associated with a bi-polar outflow along the disk axis, then the latter might be expected to show stronger absorption. Conversely, if the absorption was due to an extension of the disk itself, then the former would be expected to be stronger.

Figure \ref{Bipolar_outflow_plot} shows the radial Mg II absorption profiles for these disk galaxies for two azimuthal bins split at $|\phi| = 45\,^{\circ}$.  At small impact parameters, i.e. $b < 50$ kpc, the absorption along the disk axis is significantly stronger than in the plane of the disk.  The difference disappears at larger radii because the polar quadrants have a much steeper radial decline.  It is quite noticeable how flat the radial profile in the plane of the disk is. The errorbars in impact parameter give the standard deviation of the distribution of impact parameters within that bin. The measurements are given in Table \ref{table:sample_table3}.  

\begin{figure*}
\begin{center}
{\includegraphics[angle=0,width=12cm,height=10cm]{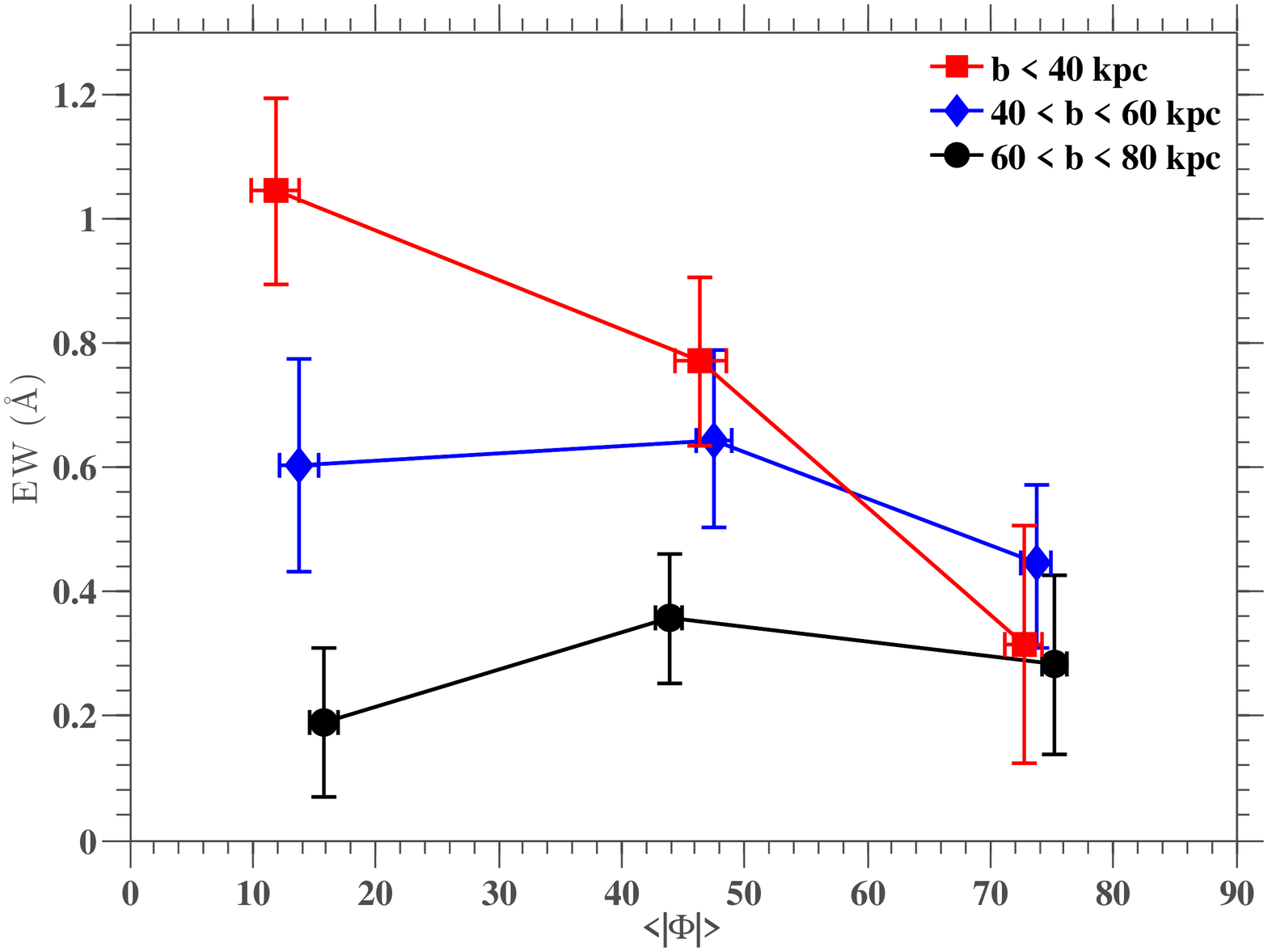}}
\caption{\label{Bipolar_outflow_plot2}Variation of Mg II EW on azimuthal angle relative to the disk axis.  Three azimuthal angular bins within $ 0\,^{\circ} < | \phi | < 30\,^{\circ} $,  $30\,^{\circ} < | \phi | < 60\,^{\circ} $ and $60\,^{\circ} < | \phi | < 90\,^{\circ} $ of the disk axis are used for each of the three bins in impact parameter, $b < 40 \;kpc$ (red square), $40 <b < 60 \;kpc$ (blue diamonds) and $60 <b < 80 \;kpc$ (black circles). The errorbars in angles are the errors on the mean of azimuthal angles within that bin.}
\end{center}
\end{figure*}

This effect is also seen in Figure \ref{Bipolar_outflow_plot2} (Table \ref{table:sample_table4}) which shows the azimuthal dependence for three radial bins.   At small impact parameters, there is a strong azimuthal gradient in absorption strength from the poles down to the plane of the disk, nominally declining by a factor of about three.   At the larger impact parameters, the effect vanishes, or may even possibly reverse (at low significance). The errorbars in azimuthal angles are the standard errors on the mean of azimuthal angles within that bin.

We take this as strong evidence that the close-in Mg II absorption within 50 kpc is associated, at least statistically, with bipolar regions aligned with the axes of symmetry of the disks. This is almost certainly the signature of a bipolar outflow of material from the disk.

\subsection {The dependence of Mg II absorption on the apparent inclination angle of disk galaxies}																													
After the submission of our original manuscript, \cite{Kacprazak2011b} have presented evidence from analysis of the inclinations of Mg II selected galaxies that the distribution of Mg II is co-planar with galaxy disks, especially for weak absorption systems.  At first sight, this is in direct contrast to the evidence in the previous section that Mg II absorption is strongest in bi-polar regions aligned with the poles of the disks. However, \citealt{Kacprazak2011b}'s result primarily refer to weaker Mg II systems at relatively large impact parameters and this may account for some of the apparent differences.

Although we believe that the azimuthal effects explored above offer a much stronger and more robust diagnostic of the geometry of the Mg II distribution than inclination effects, we present in this section, for completeness, the variation of Mg II absorption with the inclination of the associated disk galaxies.  As in the previous section, we select the foreground galaxies  to be disk-dominated, as defined in Section 2. We divide the disk galaxies into three bins in inclination angles, within $ 0\,^{\circ} < i < 50\,^{\circ} $,  $50\,^{\circ} < i < 65\,^{\circ} $ and $65\,^{\circ} < i < 90\,^{\circ} $. 

\begin{figure*}
\begin{center}
{\includegraphics[angle=0,width=12cm,height=10cm]{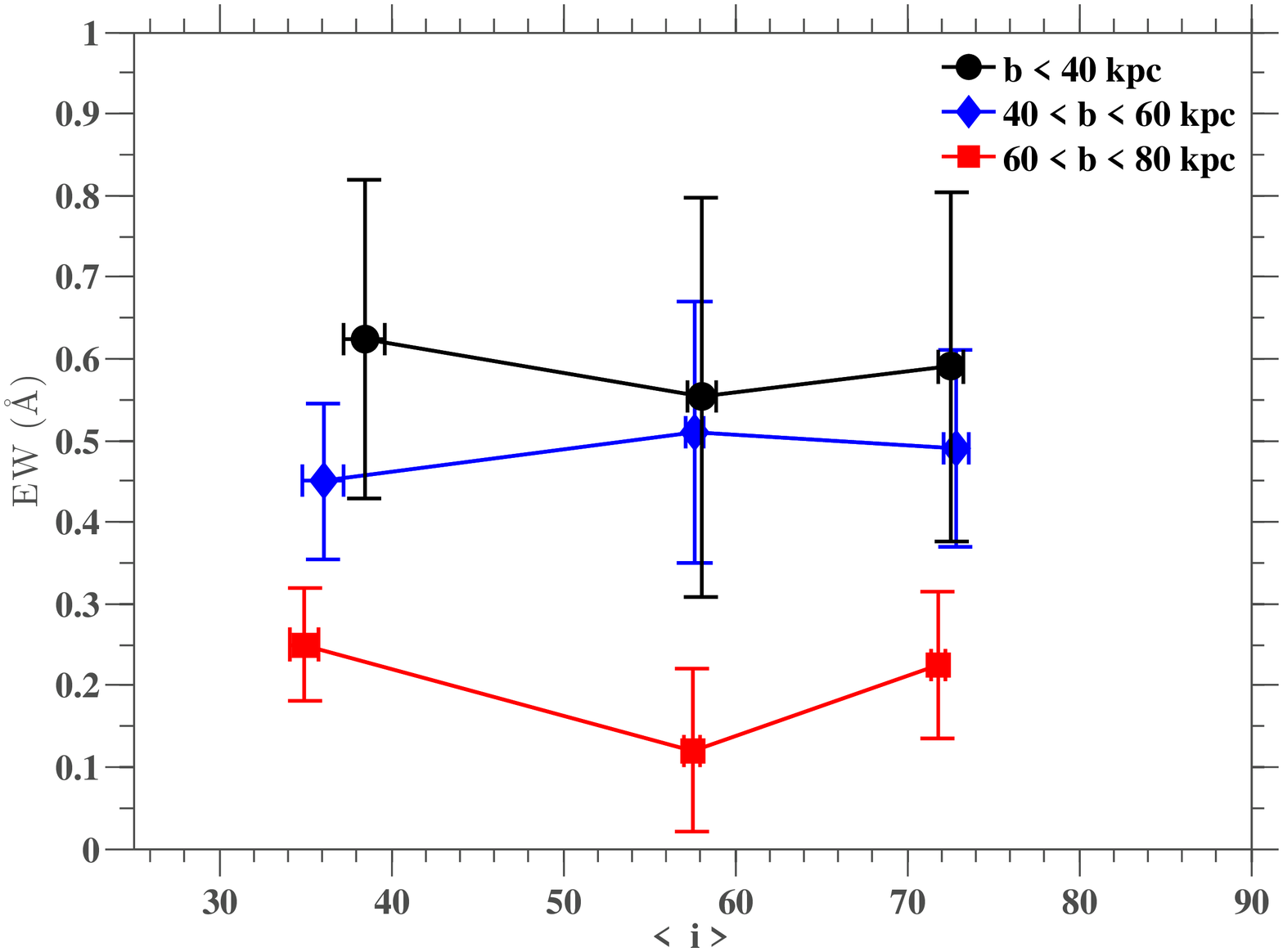}}
\caption{\label{Inclination} Variation of Mg II EW with inclination of disk-dominated galaxies. The three inclination angle bins with $ 0\,^{\circ} <  i  < 50\,^{\circ} $,  $50\,^{\circ} < i  < 65\,^{\circ} $ and $65\,^{\circ} < i < 90\,^{\circ} $ are used for each of three bins in impact parameter, $b < 40 $ kpc (black circle), $40 <b < 60$ kpc (blue diamonds) and  $60 <b < 80 $ kpc (red squares). The errorbars in angle are the errors on the mean of the inclination angles within that bin.}
\end{center}
\end{figure*}

The sense of the inclination angles is that $i = 90\,^{\circ}$ represents an edge-on system and   $i = 0\,^{\circ}$ represents a face-on system. The sample is not sub-divided in terms of azimuthal angles, i.e. for any range of inclination the derived EW are averaged over all azimuthal angles.  Figure \ref{Inclination} (Table \ref{table:sample_table5}) shows the variation of Mg II absorption around disk galaxies as a function of inclination for three impact parameter bins. The error-bars in inclination angles are the standard errors on the mean of the inclination angles within that bin. We find no significant trends of EW with the apparent inclination of disk-dominated galaxies. 

We plan to present a more comprehensive exploration of the dependence of EW on both inclination and azimuthal angles for a range of spatial distributions in a future paper. In this we will explore any differences between the two approaches described by \cite{Kacprazak2011b} and here.

\begin{deluxetable}{l c c c c }
\tablewidth{0pt}
\tabletypesize{\footnotesize}
\tablecaption{Variation of Mg II absorption with inclination around disk galaxies at different impact parameters.}
\tablehead{
\colhead{b\tablenotemark{a}} &
\colhead{$\rm{i}$\tablenotemark{b}} &
\colhead{Number\tablenotemark{c}}  &
\colhead{EW\tablenotemark{d}}  &
}
\startdata
& $65\,^{\circ} < i < 90\,^{\circ}$ & 43 & 0.59 $\pm$ 0.21 \\
 b $<$ 40 kpc & $50\,^{\circ} < i < 65\,^{\circ}$ & 39 & 0.55 $\pm$ 0.25\\
& $0\,^{\circ} < i < 50\,^{\circ}$ & 48 & 0.62 $\pm$ 0.20\\
\\
\hline
\\
& $65\,^{\circ} < i < 90\,^{\circ}$ & 55 & 0.49 $\pm$ 0.12  \\
40 $<$ b < 60 kpc & $50\,^{\circ} < i < 65\,^{\circ}$ & 54 & 0.51 $\pm$ 0.16\\
& $0\,^{\circ} < i < 50\,^{\circ}$ & 73 & 0.45 $\pm$ 0.10\\
\\
\hline
\\
& $65\,^{\circ} < i < 90\,^{\circ}$ & 96 &  0.23 $\pm$ 0.09\\
60 $<$ b < 80 kpc & $50\,^{\circ} < i < 65\,^{\circ}$ & 95 & 0.12 $\pm$ 0.10 \\
& $0\,^{\circ} < i < 50\,^{\circ}$ & 107 & 0.25 $\pm$ 0.07\\
\enddata
\tablenotetext{a}{Range of impact parameter, in physical kpc.}
\tablenotetext{b}{Range of inclination angles considered, in degrees}
\tablenotetext{c}{Number of foreground background pairs. }
\tablenotetext{d}{Measured rest frame equivalent width in \AA. }
\label{table:sample_table5}
\end{deluxetable}			

\section{Conclusions} 
In this work, we have mapped the average spatial distributions of Mg II gas around a set of zCOSMOS-bright galaxies at $0.5 < z < 0.9$, using co-added spectra from a large sample of background zCOSMOS-deep objects.  We divide the foreground galaxies in terms of their rest frame color and mass, and also by their location in or out of groups.  We also constructed average profiles for the groups around both their geometric centres and around the most massive galaxies in each group.  Finally, we examined the azimuthal dependence of Mg II absorption around inclined disk galaxies and investigated the dependence of Mg II absorption on apparent inclination of disk dominated galaxies. The main results of this study are as follows:
\begin{enumerate}

\item We find that blue foreground galaxies are associated with much stronger Mg II absorption compared to red galaxies, particularly at small impact parameters.  Within the overlapping mass range (and for $b < 50$ kpc), blue galaxies are associated with absorption systems which are about eight times stronger in equivalent width than red galaxies. This is consistent with the idea that entrained material in outflows driven by star-formation are responsible for the Mg II absorption, but could also indicate inflowing material feeding star-formation.

\item There is also a clear correlation between Mg II absorption line strength and the host galaxy stellar mass for the blue galaxies, especially at low impact parameters ($b < 50$ kpc). This may indicate a correlation with star-formation rate, since the specific star-formation rate in star-forming galaxies is known to be roughly constant.

\item For isolated (non-group) galaxies, the Mg II radial profile can be well represented by a SIS model similar to \cite{Chen2010a}. We  estimate the effective gas radius is $\rm{R_{gas}}$ = 115.2 $\pm$ 2.3 kpc (physical) and $\rm{EW_{0}}$ = 0.7 $\pm$ 0.12.  For blue galaxies it is, $\rm{R_{gas}}$ = 107.6 $\pm$ 1.3 kpc (physical) and $\rm{EW_{0}}$ = 1.1 $\pm$ 0.14, while for red galaxies it has a very similar size $\rm{R_{gas}}$ = 118.1 $\pm$ 5. kpc (physical) but much lower $\rm{EW_{0}}$ = 0.3 $\pm$ 0.11.  

\item Galaxies in groups have a much flatter absorption profile that extends beyond 140 kpc. The color dependence on Mg II line strength is maintained for the group galaxies, especially at small $b$. Likewise, the average Mg II absorption profiles constructed around a specific point in each group, either a geometric center or the location of the most massive member of the group, also are substantially more extended than the profiles around isolated galaxies.
 
\item The extended distributions of Mg II absorption in groups can however be well reproduced with a simple model that superposes the absorption profiles of individual group members, assuming each of these is represented by an absorption profile that is the same as for isolated galaxies. This might indicate that the transport of Mg II out of galaxies may not be strongly affected (within our uncertainties) by the intra-group medium, or by other processes that may occur in groups, such as tidal interactions. 

\item For disks that are seen close to edge-on (inclinations greater than 50${^\circ}$ in our convention), the Mg II absorption strength is about three times stronger in regions above the projected rotation axis of the disks (i.e. along the apparent minor axis) than in the plane of the disks (i.e. along the apparent major axis) for small impact parameters ($b < 40 $ kpc).  We suggest that this is a strong evidence that Mg II absorption is associated with bipolar regions aligned with the disk axis, presumably indicating the presence of bipolar Galactic winds. The azimuthal asymmetry is not seen at larger distances from the galaxies, indicating that the radial fall-off in absorption is faster along the disk axis than in the plane of the disk.

\item We find no statistically significant correlation between inclinations of disk galaxies and the (azimuthally averaged) Mg II absorption strength.

\end{enumerate}

\section*{Acknowledgements}
We would like to thank Nicolas Bouch{\'e} for helpful discussions on this work. We would also like to thank the anonymous referee for providing constructive and insightful comments that helped in improving the paper. This work has been supported by the Swiss National Science Foundation and is based on observations undertaken at the European Southern Observatory (ESO) Very Large Telescope (VLT) under Large Program 175.A-0839.

\bibliographystyle{thesis_bibtex}

\bibliography{mybibliography}

\begin{thebibliography}{47}
\expandafter\ifx\csname natexlab\endcsname\relax\def\natexlab#1{#1}\fi

\bibitem[{{Bergeron} \& {Stasi{\'n}ska}(1986)}]{Bergeron1986}
{Bergeron}, J. \& {Stasi{\'n}ska}, G. 1986, \aap, 169, 1

\bibitem[{{Bernet} {et~al.}(2008){Bernet}, {Miniati}, {Lilly}, {Kronberg}, \&
  {Dessauges-Zavadsky}}]{Bernet2008}
{Bernet}, M.~L., {Miniati}, F., {Lilly}, S.~J., {Kronberg}, P.~P., \&
  {Dessauges-Zavadsky}, M. 2008, \nat, 454, 302

\bibitem[{{Bolzonella} {et~al.}(2000){Bolzonella}, {Miralles}, \&
  {Pell{\'o}}}]{Hyperz}
{Bolzonella}, M., {Miralles}, J.-M., \& {Pell{\'o}}, R. 2000, AAP, 363, 476

\bibitem[{{Bolzonella} {et~al.}(2010){Bolzonella}, {Kova{\v c}}, {Pozzetti},
  {Zucca}, {Cucciati}, {Lilly}, {Peng}, {Iovino}, {Zamorani}, {Vergani},
  {Tasca}, {Lamareille}, {Oesch}, {Caputi}, {Kampczyk}, {Bardelli}, {Maier},
  {Abbas}, {Knobel}, {Scodeggio}, {Carollo}, {Contini}, {Kneib}, {Le
  F{\`e}vre}, {Mainieri}, {Renzini}, {Bongiorno}, {Coppa}, {de la Torre}, {de
  Ravel}, {Franzetti}, {Garilli}, {Le Borgne}, {Le Brun}, {Mignoli},
  {Pell{\'o}}, {Perez-Montero}, {Ricciardelli}, {Silverman}, {Tanaka},
  {Tresse}, {Bottini}, {Cappi}, {Cassata}, {Cimatti}, {Guzzo}, {Koekemoer},
  {Leauthaud}, {Maccagni}, {Marinoni}, {McCracken}, {Memeo}, {Meneux},
  {Porciani}, {Scaramella}, {Aussel}, {Capak}, {Halliday}, {Ilbert},
  {Kartaltepe}, {Salvato}, {Sanders}, {Scarlata}, {Scoville}, {Taniguchi}, \&
  {Thompson}}]{Bolzonella2010}
{Bolzonella}, M., {et~al.} 2010, \aap, 524, A76

\bibitem[{{Bouch{\'e}}(2008)}]{bouche2008}
{Bouch{\'e}}, N. 2008, \mnras, 389, L18

\bibitem[{{Bouch{\'e}} {et~al.}(2006){Bouch{\'e}}, {Murphy}, {P{\'e}roux},
  {Csabai}, \& {Wild}}]{Bouche2006}
{Bouch{\'e}}, N., {Murphy}, M.~T., {P{\'e}roux}, C., {Csabai}, I., \& {Wild},
  V. 2006, \mnras, 371, 495

\bibitem[{{Bowen} {et~al.}(1995){Bowen}, {Blades}, \& {Pettini}}]{Bowen1995}
{Bowen}, D.~V., {Blades}, J.~C., \& {Pettini}, M. 1995, \apj, 448, 662

\bibitem[{{Charlton} \& {Churchill}(1998)}]{Charlton1998}
{Charlton}, J.~C. \& {Churchill}, C.~W. 1998, \apj, 499, 181

\bibitem[{{Charlton} {et~al.}(2003){Charlton}, {Ding}, {Zonak}, {Churchill},
  {Bond}, \& {Rigby}}]{Charlton2003}
{Charlton}, J.~C., {Ding}, J., {Zonak}, S.~G., {Churchill}, C.~W., {Bond},
  N.~A., \& {Rigby}, J.~R. 2003, \apj, 589, 111

\bibitem[{{Chelouche} \& {Bowen}(2010)}]{Chelouche2010}
{Chelouche}, D. \& {Bowen}, D.~V. 2010, \apj, 722, 1821

\bibitem[{{Chen} {et~al.}(2010{\natexlab{a}}){Chen}, {Helsby}, {Gauthier},
  {Shectman}, {Thompson}, \& {Tinker}}]{Chen2010a}
{Chen}, H., {Helsby}, J.~E., {Gauthier}, J., {Shectman}, S.~A., {Thompson},
  I.~B., \& {Tinker}, J.~L. 2010{\natexlab{a}}, ApJ, 714, 1521

\bibitem[{{Chen} \& {Tinker}(2008)}]{2008C&T}
{Chen}, H. \& {Tinker}, J.~L. 2008, ApJ, 687, 745

\bibitem[{{Chen} {et~al.}(2010{\natexlab{b}}){Chen}, {Wild}, {Tinker},
  {Gauthier}, {Helsby}, {Shectman}, \& {Thompson}}]{Chen2010b}
{Chen}, H., {Wild}, V., {Tinker}, J.~L., {Gauthier}, J., {Helsby}, J.~E.,
  {Shectman}, S.~A., \& {Thompson}, I.~B. 2010{\natexlab{b}}, \apjl, 724, L176

\bibitem[{{Churchill} \& {Charlton}(1999)}]{Churchill&Charlton99}
{Churchill}, C. \& {Charlton}, J. 1999, in Bulletin of the American
  Astronomical Society, Vol.~31, American Astronomical Society Meeting
  Abstracts, 1451

\bibitem[{{Churchill} {et~al.}(2005{\natexlab{a}}){Churchill}, {Steidel}, \&
  {Kacprzak}}]{Churchill2005}
{Churchill}, C., {Steidel}, C., \& {Kacprzak}, G. 2005{\natexlab{a}}, in
  Astronomical Society of the Pacific Conference Series, Vol. 331, Extra-Planar
  Gas, ed. {R.~Braun}, 387

\bibitem[{{Churchill} {et~al.}(2005{\natexlab{b}}){Churchill}, {Kacprzak}, \&
  {Steidel}}]{Churchill2005a}
{Churchill}, C.~W., {Kacprzak}, G.~G., \& {Steidel}, C.~C. 2005{\natexlab{b}},
  Proceedings of the International Astronomical Union, 1, 24

\bibitem[{{Churchill} {et~al.}(2000){Churchill}, {Mellon}, {Charlton},
  {Jannuzi}, {Kirhakos}, {Steidel}, \& {Schneider}}]{Churchill2000}
{Churchill}, C.~W., {Mellon}, R.~R., {Charlton}, J.~C., {Jannuzi}, B.~T.,
  {Kirhakos}, S., {Steidel}, C.~C., \& {Schneider}, D.~P. 2000, \apjs, 130, 91

\bibitem[{{Ilbert} {et~al.}(2009){Ilbert}, {Capak}, {Salvato}, {Aussel},
  {McCracken}, {Sanders}, {Scoville}, {Kartaltepe}, {Arnouts}, {Floc'h},
  {Mobasher}, {Taniguchi}, {Lamareille}, {Leauthaud}, {Sasaki}, {Thompson},
  {Zamojski}, {Zamorani}, {Bardelli}, {Bolzonella}, {Bongiorno}, {Brusa},
  {Caputi}, {Carollo}, {Contini}, {Cook}, {Coppa}, {Cucciati}, {de la Torre},
  {de Ravel}, {Franzetti}, {Garilli}, {Hasinger}, {Iovino}, {Kampczyk},
  {Kneib}, {Knobel}, {Kovac}, {LeBorgne}, {LeBrun}, {F{\`e}vre}, {Lilly},
  {Looper}, {Maier}, {Mainieri}, {Mellier}, {Mignoli}, {Murayama}, {Pell{\`o}},
  {Peng}, {P{\'e}rez-Montero}, {Renzini}, {Ricciardelli}, {Schiminovich},
  {Scodeggio}, {Shioya}, {Silverman}, {Surace}, {Tanaka}, {Tasca}, {Tresse},
  {Vergani}, \& {Zucca}}]{2009ApJ...690.1236I}
{Ilbert}, O., {et~al.} 2009, ApJ, 690, 1236

\bibitem[{{Kacprzak} {et~al.}(2010{\natexlab{a}}){Kacprzak}, {Churchill},
  {Ceverino}, {Steidel}, {Klypin}, \& {Murphy}}]{Kacprzak2010a}
{Kacprzak}, G.~G., {Churchill}, C.~W., {Ceverino}, D., {Steidel}, C.~C.,
  {Klypin}, A., \& {Murphy}, M.~T. 2010{\natexlab{a}}, ApJ, 711, 533

\bibitem[{{Kacprzak} {et~al.}(2011){Kacprzak}, {Churchill}, {Evans}, {Murphy},
  \& {Steidel}}]{Kacprazak2011b}
{Kacprzak}, G.~G., {Churchill}, C.~W., {Evans}, J.~L., {Murphy}, M.~T., \&
  {Steidel}, C.~C. 2011, ArXiv e-prints:1106.3068

\bibitem[{{Kacprzak} {et~al.}(2008){Kacprzak}, {Churchill}, {Steidel}, \&
  {Murphy}}]{Kacprzak2008}
{Kacprzak}, G.~G., {Churchill}, C.~W., {Steidel}, C.~C., \& {Murphy}, M.~T.
  2008, \aj, 135, 922

\bibitem[{{Kacprzak} {et~al.}(2010{\natexlab{b}}){Kacprzak}, {Murphy}, \&
  {Churchill}}]{Kacprazak2010b}
{Kacprzak}, G.~G., {Murphy}, M.~T., \& {Churchill}, C.~W. 2010{\natexlab{b}},
  \mnras, 406, 445

\bibitem[{{Knobel} {et~al.}(2009){Knobel}, {Lilly}, {Iovino}, {Porciani},
  {Kova{\v c}}, {Cucciati}, {Finoguenov}, {Kitzbichler}, {Carollo}, {Contini},
  {Kneib}, {Le F{\`e}vre}, {Mainieri}, {Renzini}, {Scodeggio}, {Zamorani},
  {Bardelli}, {Bolzonella}, {Bongiorno}, {Caputi}, {Coppa}, {de la Torre}, {de
  Ravel}, {Franzetti}, {Garilli}, {Kampczyk}, {Lamareille}, {Le Borgne}, {Le
  Brun}, {Maier}, {Mignoli}, {Pello}, {Peng}, {Perez Montero}, {Ricciardelli},
  {Silverman}, {Tanaka}, {Tasca}, {Tresse}, {Vergani}, {Zucca}, {Abbas},
  {Bottini}, {Cappi}, {Cassata}, {Cimatti}, {Fumana}, {Guzzo}, {Koekemoer},
  {Leauthaud}, {Maccagni}, {Marinoni}, {McCracken}, {Memeo}, {Meneux}, {Oesch},
  {Pozzetti}, \& {Scaramella}}]{Knobel_2009}
{Knobel}, C., {et~al.} 2009, ApJ, 697, 1842

\bibitem[{{Koekemoer} {et~al.}(2007){Koekemoer}, {Aussel}, {Calzetti}, {Capak},
  {Giavalisco}, {Kneib}, {Leauthaud}, {Le F{\`e}vre}, {McCracken}, {Massey},
  {Mobasher}, {Rhodes}, {Scoville}, \& {Shopbell}}]{Koekemoer2007}
{Koekemoer}, A.~M., {et~al.} 2007, \apjs, 172, 196

\bibitem[{{Lanzetta} {et~al.}(1987){Lanzetta}, {Turnshek}, \&
  {Wolfe}}]{Lanzetta1987}
{Lanzetta}, K.~M., {Turnshek}, D.~A., \& {Wolfe}, A.~M. 1987, \apj, 322, 739

\bibitem[{{Lilly} {et~al.}(2007){Lilly}, {Le F{\`e}vre}, {Renzini}, {Zamorani},
  {Scodeggio}, {Contini}, {Carollo}, {Hasinger}, {Kneib}, {Iovino}, {Le Brun},
  {Maier}, {Mainieri}, {Mignoli}, {Silverman}, {Tasca}, {Bolzonella},
  {Bongiorno}, {Bottini}, {Capak}, {Caputi}, {Cimatti}, {Cucciati}, {Daddi},
  {Feldmann}, {Franzetti}, {Garilli}, {Guzzo}, {Ilbert}, {Kampczyk}, {Kovac},
  {Lamareille}, {Leauthaud}, {Borgne}, {McCracken}, {Marinoni}, {Pello},
  {Ricciardelli}, {Scarlata}, {Vergani}, {Sanders}, {Schinnerer}, {Scoville},
  {Taniguchi}, {Arnouts}, {Aussel}, {Bardelli}, {Brusa}, {Cappi}, {Ciliegi},
  {Finoguenov}, {Foucaud}, {Franceschini}, {Halliday}, {Impey}, {Knobel},
  {Koekemoer}, {Kurk}, {Maccagni}, {Maddox}, {Marano}, {Marconi}, {Meneux},
  {Mobasher}, {Moreau}, {Peacock}, {Porciani}, {Pozzetti}, {Scaramella},
  {Schiminovich}, {Shopbell}, {Smail}, {Thompson}, {Tresse}, {Vettolani},
  {Zanichelli}, \& {Zucca}}]{Lilly2007}
{Lilly}, S.~J., {et~al.} 2007, ApJS, 172, 70

\bibitem[{{Lilly} {et~al.}(2009){Lilly}, {Le Brun}, {Maier}, {Mainieri},
  {Mignoli}, {Scodeggio}, {Zamorani}, {Carollo}, {Contini}, {Kneib}, {Le
  F{\`e}vre}, {Renzini}, {Bardelli}, {Bolzonella}, {Bongiorno}, {Caputi},
  {Coppa}, {Cucciati}, {de la Torre}, {de Ravel}, {Franzetti}, {Garilli},
  {Iovino}, {Kampczyk}, {Kovac}, {Knobel}, {Lamareille}, {Le Borgne}, {Pello},
  {Peng}, {P{\'e}rez-Montero}, {Ricciardelli}, {Silverman}, {Tanaka}, {Tasca},
  {Tresse}, {Vergani}, {Zucca}, {Ilbert}, {Salvato}, {Oesch}, {Abbas},
  {Bottini}, {Capak}, {Cappi}, {Cassata}, {Cimatti}, {Elvis}, {Fumana},
  {Guzzo}, {Hasinger}, {Koekemoer}, {Leauthaud}, {Maccagni}, {Marinoni},
  {McCracken}, {Memeo}, {Meneux}, {Porciani}, {Pozzetti}, {Sanders},
  {Scaramella}, {Scarlata}, {Scoville}, {Shopbell}, \&
  {Taniguchi}}]{Lilly2009_Article}
{Lilly}, S.~J., {et~al.} 2009, ApJS, 184, 218

\bibitem[{{M{\'e}nard} {et~al.}(2009){M{\'e}nard}, {Wild}, {Nestor}, {Quider},
  \& {Zibetti}}]{Menard2009}
{M{\'e}nard}, B., {Wild}, V., {Nestor}, D., {Quider}, A., \& {Zibetti}, S.
  2009, ArXiv e-prints:- 0912.3263

\bibitem[{{Nestor} {et~al.}(2010){Nestor}, {Johnson}, {Wild}, {M{\'e}nard},
  {Turnshek}, {Rao}, \& {Pettini}}]{Nestor2010}
{Nestor}, D.~B., {Johnson}, B.~D., {Wild}, V., {M{\'e}nard}, B., {Turnshek},
  D.~A., {Rao}, S., \& {Pettini}, M. 2010, ArXiv e-prints:-1003.0693

\bibitem[{{Nestor} {et~al.}(2005){Nestor}, {Turnshek}, \& {Rao}}]{Nestor2005}
{Nestor}, D.~B., {Turnshek}, D.~A., \& {Rao}, S.~M. 2005, \apj, 628, 637

\bibitem[{{Nestor} {et~al.}(2007){Nestor}, {Turnshek}, {Rao}, \&
  {Quider}}]{Nestor2007}
{Nestor}, D.~B., {Turnshek}, D.~A., {Rao}, S.~M., \& {Quider}, A.~M. 2007,
  \apj, 658, 185

\bibitem[{{Peng} {et~al.}(2010){Peng}, {Lilly}, {Kova{\v c}}, {Bolzonella},
  {Pozzetti}, {Renzini}, {Zamorani}, {Ilbert}, {Knobel}, {Iovino}, {Maier},
  {Cucciati}, {Tasca}, {Carollo}, {Silverman}, {Kampczyk}, {de Ravel},
  {Sanders}, {Scoville}, {Contini}, {Mainieri}, {Scodeggio}, {Kneib}, {Le
  F{\`e}vre}, {Bardelli}, {Bongiorno}, {Caputi}, {Coppa}, {de la Torre},
  {Franzetti}, {Garilli}, {Lamareille}, {Le Borgne}, {Le Brun}, {Mignoli},
  {Perez Montero}, {Pello}, {Ricciardelli}, {Tanaka}, {Tresse}, {Vergani},
  {Welikala}, {Zucca}, {Oesch}, {Abbas}, {Barnes}, {Bordoloi}, {Bottini},
  {Cappi}, {Cassata}, {Cimatti}, {Fumana}, {Hasinger}, {Koekemoer},
  {Leauthaud}, {Maccagni}, {Marinoni}, {McCracken}, {Memeo}, {Meneux}, {Nair},
  {Porciani}, {Presotto}, \& {Scaramella}}]{Peng2010}
{Peng}, Y., {et~al.} 2010, ApJ, 721, 193

\bibitem[{{Petitjean} \& {Bergeron}(1990)}]{Petitjean1990}
{Petitjean}, P. \& {Bergeron}, J. 1990, \aap, 231, 309

\bibitem[{{Prochter} {et~al.}(2006){Prochter}, {Prochaska}, \&
  {Burles}}]{Prochter2006}
{Prochter}, G.~E., {Prochaska}, J.~X., \& {Burles}, S.~M. 2006, \apj, 639, 766

\bibitem[{{Rao} {et~al.}(2006){Rao}, {Turnshek}, \& {Nestor}}]{Rao2006}
{Rao}, S.~M., {Turnshek}, D.~A., \& {Nestor}, D.~B. 2006, \apj, 636, 610

\bibitem[{{Rigby} {et~al.}(2002){Rigby}, {Charlton}, \&
  {Churchill}}]{rigby2002}
{Rigby}, J.~R., {Charlton}, J.~C., \& {Churchill}, C.~W. 2002, \apj, 565, 743

\bibitem[{{Rubin} {et~al.}(2010){Rubin}, {Weiner}, {Koo}, {Martin},
  {Prochaska}, {Coil}, \& {Newman}}]{Rubin2010}
{Rubin}, K.~H.~R., {Weiner}, B.~J., {Koo}, D.~C., {Martin}, C.~L., {Prochaska},
  J.~X., {Coil}, A.~L., \& {Newman}, J.~A. 2010, \apj, 719, 1503

\bibitem[{{Sargent} {et~al.}(1988){Sargent}, {Steidel}, \&
  {Boksenberg}}]{Sargent1988}
{Sargent}, W.~L.~W., {Steidel}, C.~C., \& {Boksenberg}, A. 1988, \apj, 334, 22

\bibitem[{{Scarlata} {et~al.}(2007){Scarlata}, {Carollo}, {Lilly}, {Sargent},
  {Feldmann}, {Kampczyk}, {Porciani}, {Koekemoer}, {Scoville}, {Kneib},
  {Leauthaud}, {Massey}, {Rhodes}, {Tasca}, {Capak}, {Maier}, {McCracken},
  {Mobasher}, {Renzini}, {Taniguchi}, {Thompson}, {Sheth}, {Ajiki}, {Aussel},
  {Murayama}, {Sanders}, {Sasaki}, {Shioya}, \& {Takahashi}}]{Scarlata2007}
{Scarlata}, C., {et~al.} 2007, \apjs, 172, 406

\bibitem[{{Scoville} {et~al.}(2007){Scoville}, {Aussel}, {Brusa}, {Capak},
  {Carollo}, {Elvis}, {Giavalisco}, {Guzzo}, {Hasinger}, {Impey}, {Kneib},
  {LeFevre}, {Lilly}, {Mobasher}, {Renzini}, {Rich}, {Sanders}, {Schinnerer},
  {Schminovich}, {Shopbell}, {Taniguchi}, \& {Tyson}}]{Scoville2007}
{Scoville}, N., {et~al.} 2007, ApJS, 172, 1

\bibitem[{{Steidel}(1995)}]{Steidel1995}
{Steidel}, C.~C. 1995, in QSO Absorption Lines, ed. {G.~Meylan}, 139

\bibitem[{{Steidel} {et~al.}(2010){Steidel}, {Erb}, {Shapley}, {Pettini},
  {Reddy}, {Bogosavljevi{\'c}}, {Rudie}, \& {Rakic}}]{Steidel2010}
{Steidel}, C.~C., {Erb}, D.~K., {Shapley}, A.~E., {Pettini}, M., {Reddy}, N.,
  {Bogosavljevi{\'c}}, M., {Rudie}, G.~C., \& {Rakic}, O. 2010, \apj, 717, 289

\bibitem[{{Steidel} \& {Sargent}(1992)}]{Steidel1992}
{Steidel}, C.~C. \& {Sargent}, W.~L.~W. 1992, \apjs, 80, 1

\bibitem[{{Tinker} \& {Chen}(2008)}]{2008T&C}
{Tinker}, J.~L. \& {Chen}, H. 2008, ApJ, 679, 1218

\bibitem[{{Weiner} {et~al.}(2009){Weiner}, {Coil}, {Prochaska}, {Newman},
  {Cooper}, {Bundy}, {Conselice}, {Dutton}, {Faber}, {Koo}, {Lotz}, {Rieke}, \&
  {Rubin}}]{Wiener2009}
{Weiner}, B.~J., {et~al.} 2009, \apj, 692, 187

\bibitem[{{Zibetti} {et~al.}(2007){Zibetti}, {M{\'e}nard}, {Nestor}, {Quider},
  {Rao}, \& {Turnshek}}]{2007Zibetti}
{Zibetti}, S., {M{\'e}nard}, B., {Nestor}, D.~B., {Quider}, A.~M., {Rao},
  S.~M., \& {Turnshek}, D.~A. 2007, ApJ, 658, 161

\bibitem[{{Zucca} {et~al.}(2009){Zucca}, {Bardelli}, {Bolzonella}, {Zamorani},
  {Ilbert}, {Pozzetti}, {Mignoli}, {Kova{\v c}}, {Lilly}, {Tresse}, {Tasca},
  {Cassata}, {Halliday}, {Vergani}, {Caputi}, {Carollo}, {Contini}, {Kneib},
  {Le F{\`e}vre}, {Mainieri}, {Renzini}, {Scodeggio}, {Bongiorno}, {Coppa},
  {Cucciati}, {de La Torre}, {de Ravel}, {Franzetti}, {Garilli}, {Iovino},
  {Kampczyk}, {Knobel}, {Lamareille}, {Le Borgne}, {Le Brun}, {Maier},
  {Pell{\`o}}, {Peng}, {Perez-Montero}, {Ricciardelli}, {Silverman}, {Tanaka},
  {Abbas}, {Bottini}, {Cappi}, {Cimatti}, {Guzzo}, {Koekemoer}, {Leauthaud},
  {Maccagni}, {Marinoni}, {McCracken}, {Memeo}, {Meneux}, {Moresco}, {Oesch},
  {Porciani}, {Scaramella}, {Arnouts}, {Aussel}, {Capak}, {Kartaltepe},
  {Salvato}, {Sanders}, {Scoville}, {Taniguchi}, \& {Thompson}}]{Zucca2009}
{Zucca}, E., {et~al.} 2009, \aap, 508, 1217

\end{thebibliography}

\end{document}